\documentclass{aa}	%Optional arguments:	[letter] for a letter
\usepackage[varg]{txfonts}

\usepackage{natbib}
\bibpunct{(}{)}{;}{a}{}{,} % to follow A&A style

\usepackage{multirow}
\usepackage{lscape}

\title{Galaxy And Mass Assembly: Group and field galaxy morphologies in the star-formation rate - stellar mass plane}
\titlerunning{GAMA: Group and field galaxy morphologies in the SFR-M$_{\star}$ plane}

\author{W.~J.~Pearson\inst{\ref{inst:NCBJ}}
  \and L.~Wang\inst{\ref{inst:SRON}, \ref{inst:Kapteyn}}
  \and S.~Brough\inst{\ref{inst:NSW}}
  \and B.~W.~Holwerda\inst{\ref{inst:Louisville}}
  \and A.~M.~Hopkins\inst{\ref{inst:Macquarie}}
  \and J.~Loveday\inst{\ref{inst:Sussex}}
}

\institute{National Centre for Nuclear Research, Pasteura 7, 02-093 Warszawa, Poland\label{inst:NCBJ}\\\email{william.pearson@ncbj.gov.pl}
\and SRON Netherlands Institute for Space Research, Landleven 12, 9747 AD, Groningen, The Netherlands\label{inst:SRON}
\and Kapteyn Astronomical Institute, University of Groningen, Postbus 800, 9700 AV Groningen, The Netherlands\label{inst:Kapteyn}
\and School of Physics, University of New South Wales, NSW 2052, Australia\label{inst:NSW}
\and Department of Physics and Astronomy, 102 Natural Science Building, University of Louisville, Louisville, KY 40292, USA\label{inst:Louisville}
\and Australian Astronomical Optics, Macquarie University, 105 Delhi Road, North Ryde, NSW 2113, Australia\label{inst:Macquarie}
\and Astronomy Centre, University of Sussex, Falmer, Brighton BN1 9QH, UK\label{inst:Sussex}
}

\date{Received 13 August 2020 /
	Accepted 31 December 2020}

\abstract{}
{We study the environment in which a galaxy lies (i.e. field or group) and its connection with the morphology of the galaxy. This is done by examining the distribution of parametric and non-parametric statistics across the star-formation rate (SFR) - stellar mass (M$_{\star}$) plane and studying how these distributions change with the environment in the local universe ($z<0.15$).}
{We determine the concentration (C), Gini, M$_{20}$, asymmetry, Gini-M$_{20}$ bulge statistic (GMB), 50\% light radius ($r_{50}$), total S\'{e}rsic index, and bulge S\'{e}rsic index ($n_{Bulge}$) for galaxies from the Galaxy and Mass Assembly (GAMA) survey using optical images from the Kilo Degree Survey. We determine the galaxy environment using the GAMA group catalogue and split the galaxies into field or group galaxies. The group galaxies are further divided by the group halo mass (M$_{h}$) - $11 \leq \mathrm{log(M}_{h}/\mathrm{M}_\odot) < 12$, $12 \leq \mathrm{log(M}_{h}/\mathrm{M}_\odot) < 13$, and $13 \leq \mathrm{log(M}_{h}/\mathrm{M}_\odot) < 14$ - and into central and satellite galaxies. The galaxies in each of these samples are then placed onto the SFR-M$_{\star}$ plane, and each parameter is used as a third dimension. We fit the resulting distributions for each parameter in each sample using two two-dimensional Gaussian distributions: one for star-forming galaxies and one for quiescent galaxies. The coefficients of these Gaussian fits are then compared between environments.}
{Using C and $r_{50}$, we find that galaxies typically become larger as the group mass increases. This change is greater for larger galaxies. There is no indication that galaxies are typically more or less clumpy as the environment changes. Using GMB and $n_{Bulge}$, we see that the star-forming galaxies do not become more bulge or disk dominated as the group mass changes. Asymmetry does not appear to be greatly influenced by environment.}
{}


\begin{document}
\maketitle

\section{Introduction}\label{sec:intro}
Galaxy groups are large, gravitationally bound structures that contain a few to a few tens of galaxies, all lying within the same dark matter halo. The scale of the groups lies between those of single galaxies and those of the larger galaxy clusters: The halo masses of groups are typically between 10$^{11}$ and 10$^{14}$ M$_{\odot}$ \citep[e.g.][]{2007ApJ...655..790C, 2017MNRAS.470.2982L, 2019MNRAS.490.2367C}. Galaxy groups, and indeed clusters, arise from the mergers of smaller dark matter halos, or from the infall of a smaller dark matter halo into a larger halo, under the hierarchical growth found in the current cold dark matter cosmology. More than half of all galaxies are observed to lie in group environments, with the remaining galaxies being field galaxies that are found alone \citep[e.g.][]{1982ApJ...257..423H, 2004MNRAS.348..866E, 2017MNRAS.470.2982L}. The galaxies within groups contain a range of different morphological and physical properties, with elliptical, spiral, star-forming, and quiescent galaxies, and everything in between. Groups are also thought to pre-process galaxies before forming larger and denser structures \citep[e.g.][]{2008MNRAS.388.1152P, 2013MNRAS.432..336W, 2015ApJ...806..101H, 2018MNRAS.474..547K}.

It is often reported that there is a reduction in the star-formation rate (SFR) of galaxies, as well as the fraction of star-forming galaxies, that lie within groups compared to those that lie in the field. This reduction is seen to be greater in denser and more massive environments as well as in galaxies closer to the centre of the group \citep[e.g.][]{2002MNRAS.334..673L, 2003ApJ...584..210G, 2010ApJ...721..193P, 2012ApJ...757....4P, 2012MNRAS.423.3679W, 2013A&A...558A.100P, 2017MNRAS.464..121S, 2019MNRAS.483.2851S, 2017MNRAS.469.3670S, 2018ApJ...857...71B, 2019MNRAS.485.1528A, 2020ApJ...889..156C, 2020PASP..132e4101G, 2020MNRAS.492.2722O, 2020ApJ...898...20C, 2020arXiv200908212V}. This environmental quenching is believed to arise from harassment, galaxy-galaxy interactions and gas starvation as well as ram pressure stripping as a galaxy moves through a group \citep[e.g.][]{2010ApJ...721..193P, 2013MNRAS.432..336W, 2017MNRAS.469.3670S}. This stripping results in the star-formation becoming more centralised in star-forming galaxies that lie in higher mass groups \citep{2017MNRAS.464..121S, 2019MNRAS.483.2851S}. For the stellar mass (M$_{\star}$) of galaxies in groups, the stellar-mass function is seen to have a steeper low mass end as the environment becomes denser \citep{2016A&A...590A..29P}.

The galaxy main sequence (MS) is a well-studied, tight correlation between the SFR and M$_{\star}$ of star-forming galaxies \citep[e.g.][]{2004MNRAS.351.1151B, 2007A&A...468...33E, 2007ApJ...660L..43N, 2014ApJS..214...15S, 2018A&A...615A.146P}. With the noted reduction in SFR in group environments and changes to the low mass end of the stellar-mass function, it is possible for groups to influence the MS. However, it is as yet unclear if the MS is influenced by the environment. Some studies provide evidence that the MSs derived from galaxies within groups or lone field galaxies are consistent with one another \citep{2010ApJ...721..193P, 2018MNRAS.481.3456C, 2019A&A...625A.112G}. However, others show a reduction in the normalisation of the MS for group galaxies \citep{2018ApJ...857...71B, 2020MNRAS.493.5987O} or show that galaxies in the centre of a group have an unchanged MS, while the satellite galaxies have a reduction in the MS normalisation \citep{2018A&A...618A...1W}. This difference in the MS normalisation between group and field galaxies, or the lack there of, may be influenced by redshift \citep{2016MNRAS.455.2839E}.

The morphologies of all galaxies, both in and out of groups, can be studied in a number of ways. Morphological classification can be done visually, for example by examining images of galaxies for evidence of spiral arms and bars or asymmetries. Originally only conducted by expert astronomers, recent developments have seen the employment of citizen scientists, in projects such as Galaxy Zoo \citep{2008MNRAS.389.1179L}, and the increasing use of machine learning \citep[e.g.][]{2015MNRAS.450.1441D, 2015ApJS..221....8H, 2018MNRAS.476.5516B} has allowed visual morphologies to be derived for larger and larger data sets. Automated methods can also fit light profiles to galaxies, such as the S\'{e}rsic profile \citep{1963BAAA....6...41S, 2005PASA...22..118G}, to determine parametric statistics, or then can use non-parametric statistics, such as concentration, asymmetry, or smoothness \citep[e.g.][]{2000AJ....119.2645B, 2003AJ....126.1183C, 2003ApJS..147....1C, 2004AJ....128..163L}, to describe the light profile of the galaxy. From these parametric and non-parametric statistics, the Hubble type of a galaxy can be inferred or evidence of galaxy-galaxy interactions can be found.

Group environments are known to influence the morphologies of galaxies \citep{1980ApJ...236..351D}. As the mass of a group's halo increases, the fraction of spiral galaxies decreases and the fraction of elliptical galaxies increases. This trend also holds true in the field, with a greater fraction of spiral galaxies and a lower fraction of ellipticals relative to groups. The exact split between spirals and ellipticals varies between studies, but it is typically found that over half of group galaxies are elliptical, compared to less than 40\% of field galaxies \citep[e.g.][]{2000ApJ...541...95V, 2007ApJ...670..190H, 2007ApJ...670..206V, 2012ApJ...746..160W, 2013A&A...555A...5N, 2018MNRAS.481.3456C, 2020ApJ...898...20C}. As noted above, this increase in the fraction of elliptical galaxies in group environments is accompanied by environmental quenching and the associated reduction in SFR.

In this work, we aim to study the connection between galaxy morphology and the environment in which the galaxy lies: field or group. This will be done by examining the distribution of parametric and non-parametric statistics across the SFR-M$_{\star}$ plane for star-forming and quiescent galaxies and by studying how these distributions change with environment. To achieve this, we will use galaxies from the Galaxy And Mass Assembly (GAMA) survey \citep{2011MNRAS.413..971D}, which provides us with the necessary SFRs, M$_{star}$s, and environments, coupled with morphological parameters derived from the Kilo Degree Survey \citep[KiDS;][]{2013Msngr.154...44D, 2013ExA....35...25D} optical imaging in \citet{2019A&A...631A..51P}.

The paper is structured as follows. Section \ref{sec:data} discusses the data used and the sample selection, with Sect. \ref{sec:params} providing descriptions of the morphological parameters used and the modelling of their distributions. Section 4 presents the results, and Sect. 5 provides a discussion. We conclude in Sect. 6. Where necessary, we use the Planck 2015 cosmology: $\Omega_{m}$ = 0.307, $\Omega_{\Lambda}$ = 0.693 and $H_{0}$ = 67.7 \citep{2016A&A...594A..13P}

\section{Data}\label{sec:data}
This work is based on data from the GAMA survey \citep{2011MNRAS.413..971D}, using galaxies with spectroscopic redshifts below 0.15. The GAMA group catalogue \citep{2011MNRAS.416.2640R} is used to identify the galaxies that can be considered to lie in a group or those that are lone field galaxies. The group galaxies are subsequently divided by the halo mass (M$_{h}$) estimate of the group: halo mass of $11 \leq \mathrm{log(M}_{h}/\mathrm{M}_\odot) < 12$ (hereafter HB11), $12 \leq \mathrm{log(M}_{h}/\mathrm{M}_\odot) < 13$ (hereafter HB12), and $13 \leq \mathrm{log(M}_{h}/\mathrm{M}_\odot) < 14$ (hereafter HB13). The median unbiased halo mass estimate is used for M$_{h}$ \citep[See][for further details]{2011MNRAS.416.2640R}. As it is expected that central and satellite galaxies of a group will be influenced differently, we further split the central galaxies from their satellite counterparts within each halo mass bin, using the iteratively derived classifications from the GAMA group catalogue\footnote{The iterative process finds the centre of light in the r-band, removes the furthest galaxy and then repeats this process until only two galaxies remain. The central is then defined as the brightest of these two galaxies in the r-band \citep{2011MNRAS.416.2640R}.}. Larger and smaller halo masses were not considered due to the small number of galaxies available to study. We use the GAMA SFRs and M$_{\star}$s derived using MAGPHYS \citep{2008MNRAS.388.1595D}. We note that the GAMA MAGPHYS SFRs are consistent with the GAMA H$\alpha$ SFRs while providing a larger sample \citep{2016MNRAS.461..458D}.

Mass completeness was then derived empirically following \citet{2010A&A...523A..13P} for the field, central HB11, satellite HB11, central HB12, satellite HB12, central HB13, and satellite HB13 galaxies separately. To do this, the mass that every galaxy would need (M$_{lim}$) to be detected at the r-band magnitude limit was calculated for each galaxy individually following
\begin{equation}
	\mathrm{log(M}_{lim}/\mathrm{M}_{\odot}) = \mathrm{log(M/M_{\odot}}) - 0.4(r_{lim} - r),
\end{equation}
where $r$ is the observed r-band magnitude and $r_{lim}$ is the limiting magnitude used to select galaxies in GAMA: 19.8. The mass limit is then the M$_{lim}$ that 90\% of the faintest 20\% of galaxies lie below. Galaxies that lie below these mass limits are removed from further study. Once cut for completeness, the field contains 17\,678 galaxies, while the central HB11 and satellite HB11 contain 1018 and 1296 galaxies, respectively, the central and satellite HB12 contain 1780 and 3299 galaxies, and the central and satellite HB13 contain 900 and 3665 galaxies.

\begin{figure}
	\resizebox{\hsize}{!}{\includegraphics{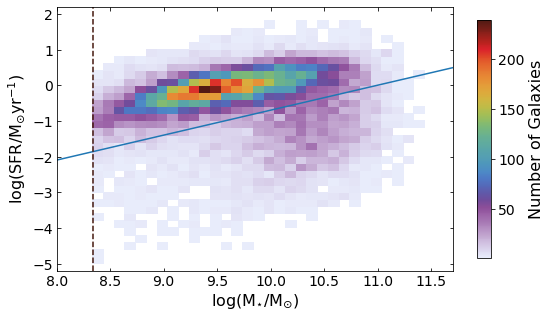}}
	\caption{SFR against M$_{\star}$ for the field galaxies showing the number density from low (light purple) to high (red). The blue line indicates the visually defined cut between the star-forming and quiescent galaxies as defined by Eq. \ref{eq:sfcut}, while the vertical dashed brown line indicates the mass limit.}
	\label{fig:SF-cut}
\end{figure}

For further analysis, we split the galaxies into star-forming and quiescent subsets. To do this, we applied a cut in the SFR-M$_{\star}$ plane. Galaxies that lie above the visually defined cut of
\begin{equation}\label{eq:sfcut}
	\mathrm{log(SFR/M}_{\odot}\mathrm{yr}^{-1}) = 0.7 \times \mathrm{log(M}_{\star}/\mathrm{M}_{\odot}) - 7.7
\end{equation}
were considered to be star-forming and those below quiescent. This line was selected to lie approximately along the green valley of the field galaxies, between the star-forming and quiescent galaxies, as shown in Fig. \ref{fig:SF-cut}.

The majority of the morphological parameters of the GAMA objects were derived in \citet{2019A&A...631A..51P} using r-band images from KiDS \citep{2013Msngr.154...44D, 2013ExA....35...25D, 2019A&A...625A...2K}, which mainly trace the stellar emission from galaxies, and the \texttt{statmorph} package \citep{2019MNRAS.483.4140R}. Here we use concentration (C), asymmetry (A), Gini, the second-order moment of the brightest 20\% of the light (M$_{20}$), and the Gini-M$_{20}$ bulge (GMB) non-parametric statistics \citep{2000AJ....119.2645B, 2001defi.conf..170W, 2003AJ....126.1183C, 2003ApJS..147....1C, 2004AJ....128..163L, 2015MNRAS.454.1886S, 2019MNRAS.483.4140R} as well as the radius containing 50\% of the light ($r_{50}$, in kpc) and the total S\'{e}rsic index \citep[$n$;][]{1963BAAA....6...41S}. These parameters are described in more detail in the next section.

The bulge S\'{e}rsic index ($n_{Bulge}$) was also studied. Here, the values from \citet{2017MNRAS.466.1513R} were used, derived using bulge-disk decomposition on r-band KiDS images with \texttt{PROFIT} \citep{2017MNRAS.466.1513R}. For the galaxies that did not have $n_{Bulge}$ values available in this catalogue, primarily galaxies with $z > 0.07$, bulge-disk decomposition was performed with \texttt{PROFIT} on the r-band KiDS images. This was done by fitting a S\'{e}rsic profile along with an exponential disk. The bulge-disk decomposition was considered successful if the centre of the exponential disk was within one-fifth of the S\'{e}rsic radius (in pixels) or three pixels of the centre of the S\'{e}rsic profile, whichever was larger. The centre of the S\'{e}rsic profile must also be further than 10~pixels from the edge of the galaxy cutout. We also require that the magnitude of the S\'{e}rsic and exponential profiles are less than 26. As a result of these criteria for successful bulge-disk decomposition, the number of galaxies in each data set is reduces when studying $n_{Bulge}$. There are 5653 field galaxies, 414 central and 497 satellite HB11 galaxies, 609 central and 1155 satellite HB12 galaxies and 289 central and 1715 satellite galaxies in HB13.

\section{Parameter descriptions and modelling}\label{sec:params}
\subsection{Parameter descriptions}
The concentration is a description of the ratio between the amount of light in the centre of a galaxy with the amount of light across a larger radius. The \texttt{statmorph} package follows the definition of \citet{2004AJ....128..163L}, comparing the ratio of the radius that contains 20\% of the light to that which contains 80\% of the light. Larger values of C indicate that more light is concentrated in the centre of the galaxy.

The asymmetry calculation also follows \citet{2004AJ....128..163L} and measures the rotational symmetry of a galaxy. To calculate A, the image is rotated by 180$^{\circ}$ and this rotated image is subtracted from the original image. The residual values are summed to give the final value of A. Larger values of A indicate that a galaxy is less rotationally symmetric.

The Gini coefficient describes the distribution of light between pixels, where a Gini of 1 has all the light in a single pixel and a value of 0 has the light spread equally across all pixels. While similar to C, Gini provides an indication of how concentrated the light is within a galaxy independent of the spatial distribution of the light within the galaxy. Gini is calculated by determining the mean of the absolute difference between all pixels, as described in \citet{2004AJ....128..163L}.

The second-order moment of the brightest 20\% of the light (M$_{20}$), as with the parameters discussed in detail so far, also follows \citet{2004AJ....128..163L}. M$_{20}$ describes the second-order moment of the brightest 20\% of a galaxy's pixels normalised by the second-order moment of the entire galaxy. To determine the second-order moment, the flux in each pixel being used is multiplied by the distance to the centre of the galaxy, with the sum of these values giving the second-order moment. Less negative M$_{20}$ values imply that a galaxy is more concentrated although, as with Gini, this concentration is not necessarily in the centre of the galaxy.

The Gini-M$_{20}$ bulge parameter (GMB) is five times the perpendicular distance, in the Gini-M$_{20}$ plane, from a galaxy to the line that separates early and late type galaxies. That is \citep{2019MNRAS.483.4140R},
\begin{equation}
	GMB = -0.693 M_{20} + 4.95 Gini - 3.96.
\end{equation}
The greater GMB is above zero, the greater the bulge domination while the lower GMB is below zero, the greater the disk domination. While providing similar information to M$_{20}$, C or the S\'{e}rsic index, GMB is less sensitive to dust and mergers \citep{2015MNRAS.454.1886S}.

The radius that contains 50\% of the light ($r_{50}$) is the radius of a circle that contains 50\% of the total light emitted from a galaxy. Here, the total light is defined as the sum of the flux within 1.5 times the circular pretrosian radius. The larger $r_{50}$ is at a given total light emitted by the galaxy, the more diffuse the light profile.

The total S\'{e}rsic index ($n$) is the best fit power law index for the S\'{e}rsic profile \citep{1963BAAA....6...41S} that has been fitted to the light profile of an entire galaxy. Higher $n$ imply a more bulge dominated structure while lower $n$ imply a more disk-like structure. If the total S\'{e}rsic profile is a good description of the light profile of a galaxy, $n$ will be monotonically related to C \citep{2001AJ....122.1707G}.

The final parameter, the bulge S\'{e}rsic index ($n_{Bulge}$), is as $n$ but for only the bulge of a galaxy after bulge-disk decomposition. Here, we assume an exponential disk when decomposing a galaxy.

\subsection{Parameter distribution modelling}
To compare the parameters in different environments, we generate simple models to describe their distributions. Two-dimensional (i.e. $z$ as a function of $x$ and $y$) Gaussian distributions are fitted in the three-dimensional SFR-M$_{\star}$-Parameter space using Markov chain Monte Carlo, implemented with the Python \texttt{emcee} package \citep{2013PASP..125..306F}. We separately fit one Gaussian for the star-forming population and one for the quiescent population. The Gaussian takes the form
\begin{equation}\label{eq:single}
\begin{split}
	z =  A_{0} + A_{G}~\exp\bigg( \frac{-1}{2(1-\rho^{2})} \Bigg( \frac{(x-\mu_x)^2}{\sigma_{x}^{2}} - & \frac{2\rho(x-\mu_x)(y-\mu_y)}{\sigma_{x}\sigma_{y}} +\\
	&\frac{(y-\mu_y)^2}{\sigma_{y}^{2}} \Bigg) \bigg),
\end{split}
\end{equation}
where $z$ is the parameter at a log(SFR)-log(M$_{\star}$) (y, x) position, $A_{0}$ is the normalisation, $A_{G}$ is the amplitude of the Gaussian, $\sigma_{x}$ is the standard deviation along log(M$_{\star}$) (hereafter $\sigma_{M_{\star}}$), $\sigma_{y}$ is the standard deviation along log(SFR) (hereafter $\sigma_{SFR}$), and $\rho$ is the correlation between log(M$_{\star}$) and log(SFR). The two means, $\mu_{x}$ and $\mu_{y}$, are the means of the distributions along log(M$_{\star}$) and log(SFR) (hereafter $\mu_{M_{\star}}$ and $\mu_{SFR}$, respectively). The absolute value of $A_{G}$ can be taken as the range of values that the parameter takes, while $A_{0}$ is the minimum value of the parameter if $A_{G}$ is positive or the maximum value if $A_{G}$ is negative. We enforce that $\sigma_{M_{\star}}$ and $\sigma_{SFR}$ must be positive and $\rho$ must be between $-1$ and $1$. Modelling in such a way allows us to study how the parameter changes with M$_{\star}$ and SFR simultaneously.

\begin{figure}
	\resizebox{\hsize}{!}{\includegraphics{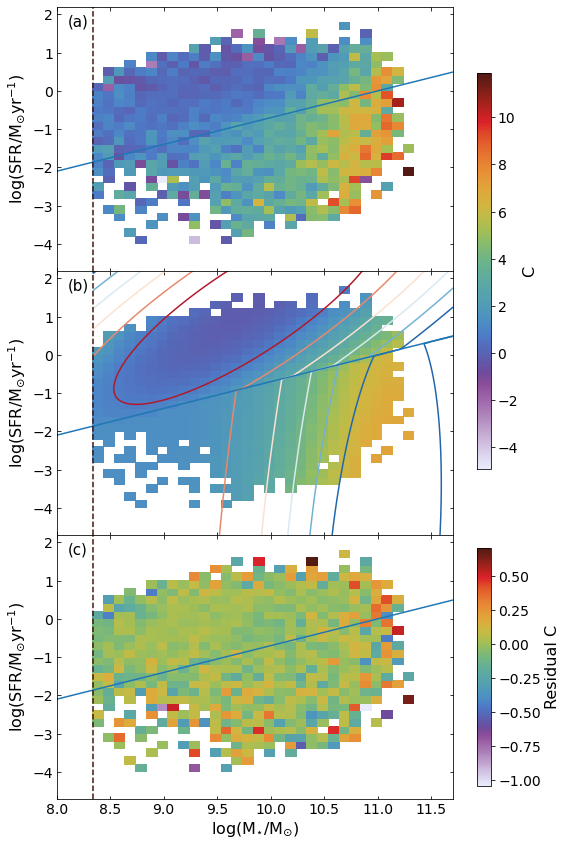}}
	\caption{Example model fitting for the field galaxies' concentration. A plot of SFR against M$_{\star}$ for the field galaxies, colour coded by the average concentration within each SFR-M$_{\star}$ bin. The bins containing at least 50 galaxies are shown. Panel (a) is the observed data, panel (b) is the model, and (c) is the residual from subtracting the model from the observations while the blue line indicates the split used for star-forming and quiescent galaxies (Eq. \ref{eq:sfcut}) and the vertical dashed brown line indicates the mass limit. Panels (a) and (b) share a colour scale. Contours for the model are overlaid in panel (b) from low values (red) to high (blue).}
	\label{fig:example-MS}
\end{figure}

An example of the fitting procedure for the field galaxies' C is presented in Fig. \ref{fig:example-MS}, where we show the mean C within bins of SFR and M$_{\star}$. As can be seen, the model (Fig. \ref{fig:example-MS}b) closely resembles the original distribution (Fig. \ref{fig:example-MS}a) with small residual values (Fig. \ref{fig:example-MS}c). As Eq. \ref{eq:sfcut} passes through SFR-M$_{\star}$ bins, these bins will contain galaxies classified as star-forming and quiescent. This results in these bins smoothing the join between the star-forming and quiescent models, hiding any discontinuities. These discontinuities can be more clearly seen in the contours of the model overlaid on Fig. \ref{fig:example-MS}b. An example corner plot showing the correlation between the fitting coefficients for the star-forming field galaxies' concentration can be found in Appendix \ref{app:corner}.

\section{Results}\label{sec:results}
\begin{figure}
	\resizebox{\hsize}{!}{\includegraphics{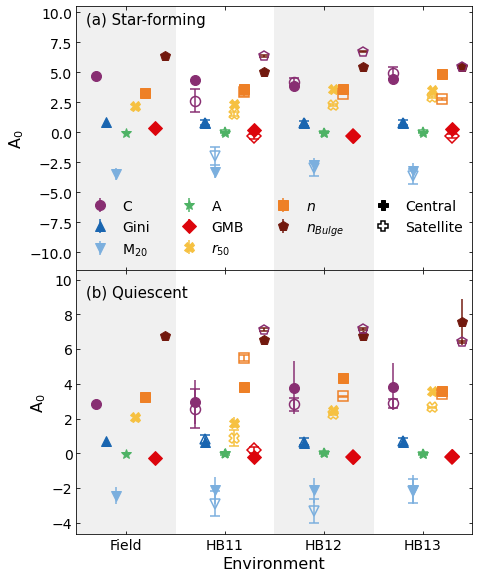}}
	\caption{Normalisations ($A_{0}$) as a function of environment for concentration (C, purple circles), Gini (dark blue upward pointing triangles), M$_{20}$ (light blue downward pointing triangles), asymmetry (A, green stars), Gini-M$_{20}$ bulge (GMB, red diamonds), $r_{50}$ (yellow crosses), total S\'{e}rsic index ($n$, orange squares), and bulge S\'{e}rsic index ($n_{Bulge}$, brown pentagrams). For the group environments, central galaxies are solid markers and satellite galaxies are empty markers. Panels (a) and (b) show the coefficients for the star-forming and quiescent galaxies, as defined in Eq \ref{eq:sfcut}, respectively.}
	\label{fig:A0}
\end{figure}

\begin{figure}
	\resizebox{\hsize}{!}{\includegraphics{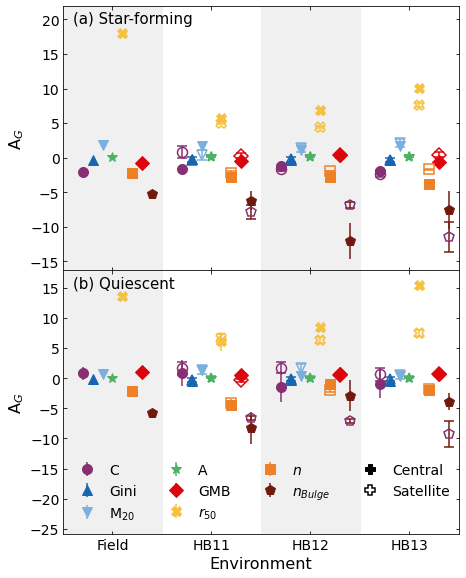}}
	\caption{Amplitudes of the Gaussian ($A_{G}$) as a function of environment for concentration (C, purple circles), Gini (dark blue upward pointing triangles), M$_{20}$ (light blue downward pointing triangles), asymmetry (A, green stars), Gini-M$_{20}$ bulge (GMB, red diamonds), $r_{50}$ (yellow crosses), total S\'{e}rsic index ($n$, orange squares), and bulge S\'{e}rsic ($n_{Bulge}$, brown pentagrams). For the group environments, central galaxies are solid markers and satellite galaxies are empty markers. Panels (a) and (b) show the coefficients for the star-forming and quiescent galaxies, as defined in Eq \ref{eq:sfcut}, respectively.}
	\label{fig:AG}
\end{figure}

\begin{figure}
	\resizebox{\hsize}{!}{\includegraphics{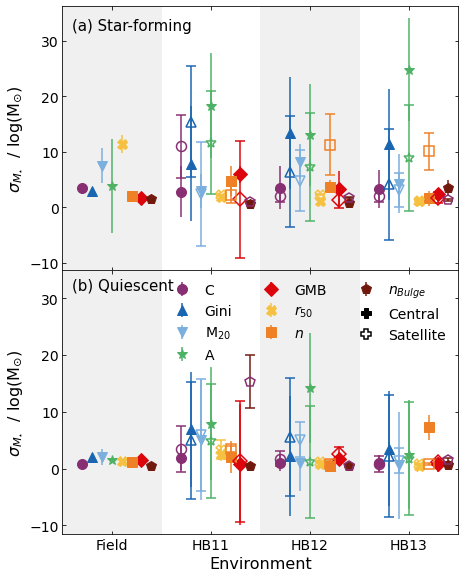}}
	\caption{$\sigma_{M_{\star}}$ coefficient, the standard deviation along M$_{\star}$, from the Gaussian fitting as a function of environment for concentration (C, purple circles), Gini (dark blue upward pointing triangles), M$_{20}$ (light blue downward pointing triangles), asymmetry (A, green stars), Gini-M$_{20}$ bulge (GMB, red diamonds), $r_{50}$ (yellow crosses), total S\'{e}rsic index ($n$, orange squares), and bulge S\'{e}rsic index ($n_{Bulge}$, brown pentagrams). For the group environments, central galaxies are solid markers and satellite galaxies are empty markers. Panels (a) and (b) show the coefficients for the star-forming and quiescent galaxies, as defined in Eq \ref{eq:sfcut}, respectively.}
	\label{fig:a}
\end{figure}

\begin{figure}
	\resizebox{\hsize}{!}{\includegraphics{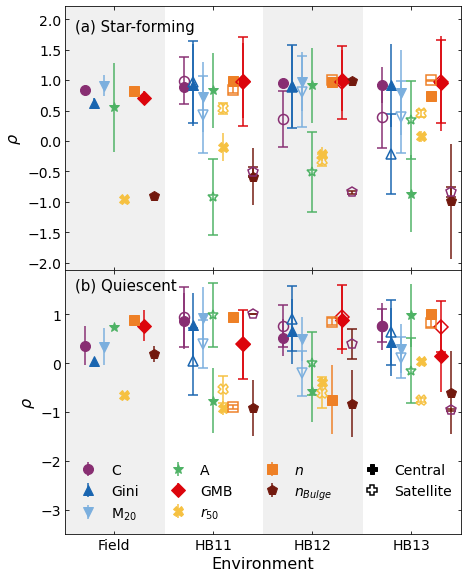}}
	\caption{$\rho$ coefficient, the correlation between M$_{\star}$ and SFR, from the Gaussian fitting as a function of environment for concentration (C, purple circles), Gini (dark blue upward pointing triangles), M$_{20}$ (light blue downward pointing triangles), asymmetry (A, green stars), Gini-M$_{20}$ bulge (GMB, red diamonds), $r_{50}$ (yellow crosses), total S\'{e}rsic index ($n$, orange squares), and bulge S\'{e}rsic index ($n_{Bulge}$, brown pentagrams). For the group environments, central galaxies are solid markers and satellite galaxies are empty markers. Panels (a) and (b) show the coefficients for the star-forming and quiescent galaxies, as defined in Eq \ref{eq:sfcut}, respectively.}
	\label{fig:b}
\end{figure}

\begin{figure}
	\resizebox{\hsize}{!}{\includegraphics{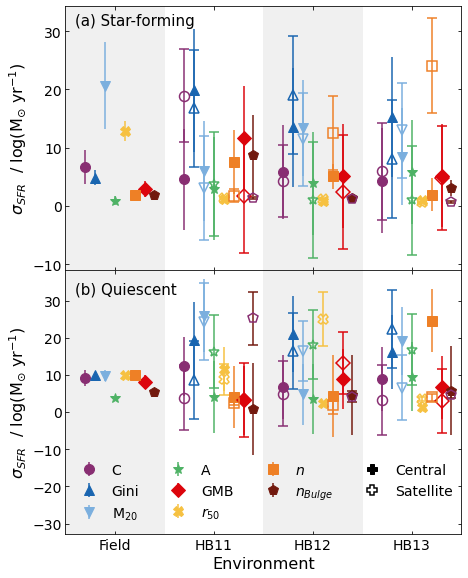}}
	\caption{$\sigma_{SFR}$ coefficient, the standard deviation along SFR, from the Gaussian fitting as a function of environment for concentration (C, purple circles), Gini (dark blue upward pointing triangles), M$_{20}$ (light blue downward pointing triangles), asymmetry (A, green stars), Gini-M$_{20}$ bulge (GMB, red diamonds), $r_{50}$ (yellow crosses), total S\'{e}rsic index ($n$, orange squares), and bulge S\'{e}rsic index ($n_{Bulge}$, brown pentagrams). For the group environments, central galaxies are solid markers and satellite galaxies are empty markers. Panels (a) and (b) show the coefficients for the star-forming and quiescent galaxies, as defined in Eq \ref{eq:sfcut}, respectively.}
	\label{fig:c}
\end{figure}

\begin{figure}
	\resizebox{\hsize}{!}{\includegraphics{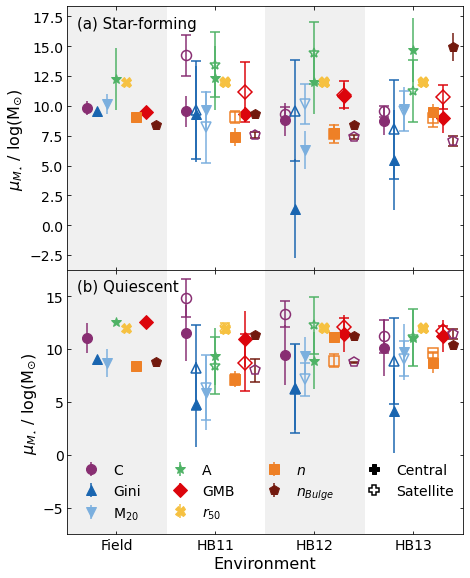}}
	\caption{$\mu_{M_{\star}}$ coefficient, the mean along M$_{\star}$, from the Gaussian fitting as a function of environment for concentration (C, purple circles), Gini (dark blue upward pointing triangles), M$_{20}$ (light blue downward pointing triangles), asymmetry (A, green stars), Gini-M$_{20}$ bulge (GMB, red diamonds), $r_{50}$ (yellow crosses), total S\'{e}rsic index ($n$, orange squares), and bulge S\'{e}rsic index ($n_{Bulge}$, brown pentagrams). For the group environments, central galaxies are solid markers and satellite galaxies are empty markers. Panels (a) and (b) show the coefficients for the star-forming and quiescent galaxies, as defined in Eq \ref{eq:sfcut}, respectively.}
	\label{fig:mu-x}
\end{figure}

\begin{figure}
	\resizebox{\hsize}{!}{\includegraphics{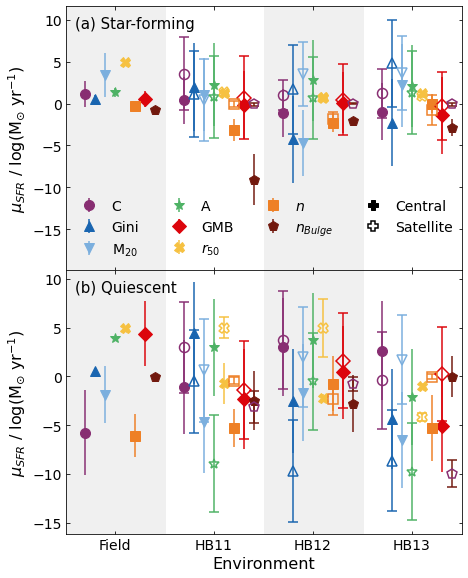}}
	\caption{$\mu_{SFR}$ coefficient, the mean along SFR, from the Gaussian fitting as a function of environment for concentration (C, purple circles), Gini (dark blue upward pointing triangles), M$_{20}$ (light blue downward pointing triangles), asymmetry (A, green stars), Gini-M$_{20}$ bulge (GMB, red diamonds), $r_{50}$ (yellow crosses), total S\'{e}rsic index ($n$, orange squares), and bulge S\'{e}rsic index ($n_{Bulge}$, brown pentagrams). For the group environments, central galaxies are solid markers and satellite galaxies are empty markers. Panels (a) and (b) show the coefficients for the star-forming and quiescent galaxies, as defined in Eq \ref{eq:sfcut}, respectively.}
	\label{fig:mu-y}
\end{figure}

Here we present the results from the fitting. The Gaussian's coefficients are presented as a function of environment mass in Figs \ref{fig:A0} to \ref{fig:mu-y} with the coefficients for star-forming galaxies in the (a) panels and for quiescent galaxies in the (b) panels. The exact values for these coefficients can be found in Appendix \ref{app:coefficients}.

Using our crude split between star-forming and quiescent galaxies, we also find that the fraction of star-forming galaxies decreases as the environment becomes more massive: 79.1\% of field galaxies are star-forming, dropping to 72.6\% in HB11, 62.0\% in HB12, and 48.7\% in the most massive environments of HB13. This is qualitatively in line with other works, which also find an increase in the fraction of quiescent galaxies as the host environment's mass increases \citep[e.g.][]{2002MNRAS.334..673L, 2003ApJ...584..210G, 2010ApJ...721..193P, 2012ApJ...757....4P, 2012MNRAS.423.3679W, 2013A&A...558A.100P, 2017MNRAS.469.3670S, 2019MNRAS.485.1528A, 2020ApJ...889..156C, 2020PASP..132e4101G, 2020MNRAS.492.2722O, 2020ApJ...898...20C}. If we subdivide the group environments further, into central and satellite galaxies, we find that a smaller percentage of central galaxies are star-forming when compared to the satellite galaxies. For HB11, HB12, and HB13 we find that 69.2\%, 51.2\%, and 40.1\% of central galaxies are star-forming, respectively, compared 75.4\%, 68.0\%, and 51.1\% of satellite galaxies. A lower star-forming fraction of central galaxies has been seen in other works, along with the decrease in star-forming central and satellite galaxies separately as the group halo mass increases \citep[e.g.][]{2017MNRAS.469.3670S, 2020MNRAS.492.2722O}.

\subsection{Concentration}
For concentration, we find that the standard deviation in M$_{\star}$ ($\sigma_{M_{\star}}$) for the star-forming central galaxies in HB11, HB12, and HB13 is consistent with $\sigma_{M_{\star}}$ of the field galaxies. For the star-forming satellite galaxies, HB11 has a $\sigma_{M_{\star}}$ that is larger the field, and slightly lower $\sigma_{M_{\star}}$ in HB12, which is also consistent with HB13, as can be seen in Fig. \ref{fig:a}a (purple circles). For the quiescent galaxies, $\sigma_{M_{\star}}$ remains consistent with the field for the central galaxies in all environments, but shows a weak indication of reducing as the group mass increases, while the satellite galaxies show a much stronger indication of reducing as the halo mass increases but again remains within error of the field in all three group mass bins.

For the standard deviation along SFR ($\sigma_{SFR}$), the star-forming satellite galaxies show a slight decreasing trend as the halo mass increases, starting higher than the field galaxies in HB11 and falling for HB12 and HB13, although the last two environments are within error of the field galaxies' value. The star-forming central galaxies are consistent with the field in HB11, HB12 and HB13. For the quiescent galaxies, the field and central galaxies are again all consistent with one another. The satellite galaxies have a lower $\sigma_{SFR}$ than the field but due to the large uncertainties are consistent with the field, as can be seen in the lower panel of Fig. \ref{fig:c}.

Thus, the rate of change of concentration as M$_{\star}$ changes is not greatly influenced by the environment in which a galaxy lies for both the star-forming and quiescent central galaxies. The C with respect to M$_{\star}$ of satellite galaxies, however, are more easily influenced by their environment with an apparent slight decrease in $\sigma_{M_{\star}}$ as the group halo mass increases. The rate of change of concentration as SFR changes is less variable compared to M$_{\star}$, with less clear reductions in $\sigma_{SFR}$ as the halo mass increases. However, due to the large uncertainties, only the satellite HB11 galaxies are inconsistent with the field value for both M$_{\star}$ and SFR.

For the correlation between SFR and M$_{\star}$ ($\rho$), we find that $\rho$ is higher than the value in the field in group environments for the star-forming central galaxies and lower for the star-forming satellite galaxies, except the satellite HB11 galaxies. For the quiescent galaxies, nearly all group environments for the central and satellite galaxies are consistent with the field's $\rho$, but the values they take are larger, as can be seen in Fig. \ref{fig:b}b.

The absolute values of $A_{G}$ for the star-forming galaxies decreases as the environment becomes more massive out to HB12, before increasing again in HB13 for the central galaxies while the satellite galaxies increase from HB11 to HB13, as shown in Fig. \ref{fig:AG}a. This implies that the range of C decreases as the host halo mass increases to HB12 before increasing again in HB13. $A_{G}$ is negative for all but the satellite galaxies in HB11, thus the normalisations, $A_{0}$, can be considered the maximum value of C and $A_{0}+A_{G}$ the minimum for all samples except the satellite HB11, where $A_{0}$ is the minimum value of C and $A_{0}+A_{G}$ the maximum. For the central galaxies, $A_{0}$ shows a slight decline from the field to HB12 before a slight increase in HB13, similar to the trend in the absolute $A_{G}$. For the satellite galaxies, there is a large drop from the field to HB11 before $A_{0}$ increases as the environment becomes more massive. This apparent large drop for HB11 is a result of $A_{G}$ being positive for the satellite galaxies in this halo-mass bin. The minimum value of C for the star-forming central galaxies, $A_{0}+A_{G}$, remains constant across all environments. $A_{0}+A_{G}$ also remains constant in all environments for the star-forming satellite galaxies, resulting in the minimum values of HB11 and HB13 being consistent with the field.

For the quiescent galaxies, the absolute values of $A_{G}$ are consistent, within error, across the majority of environments and for the central and satellite galaxies, the exception being the satellite HB12 galaxies. This suggests that the range of C is constant in all environments. All but the $A_{G}$ for the central HB12 and HB13 are positive, meaning that $A_{0}$ describes the minimum value, and $A_{0}+A_{G}$ the maximum value, for C in all but the central HB12 and HB13 samples. The $A_{0}$ values for all samples except the satellite HB12 remain consistent with the field, implying that there is no change to the minimum value of C as the environment changes for both the satellite and central galaxies, as is shown in Fig. \ref{fig:A0}. For the $A_{0}+A_{G}$ values, these are again consistent with the field, within error, in all samples. The $A_{0}+A_{G}$ in the central HB12 and HB13 samples are lower than the other central samples and is more in line with the $A_{0}$ of the central HB11 samples. This is expected as for the central HB12 and HB13 as $A_{0}+A_{G}$ describes the minimum value.

The means along M$_{\star}$ ($\mu_{M_{\star}}$) for the central and satellite star-forming galaxies are approximately consistent with each other in all group halo mass bins and are consistent with the $\mu_{M_{\star}}$ in the field, as can be seen in Fig. \ref{fig:mu-x}a. $\mu_{M_{\star}}$ is at a higher M$_{\star}$ for the satellite HB11 galaxies than the other satellite star-forming galaxies, suggesting that the peak of the distribution of C is at higher M$_{\star}$. For the quiescent galaxies, the central galaxies are again consistent with each other. The quiescent satellite galaxies show $\mu_{M_{\star}}$ moving to lower M$_{\star}$ as the halo mass increases, starting at higher M$_{\star}$ than the field and becoming consistent with the field in HB13.

The mean along SFR ($\mu_{SFR}$) show a slight decreasing trend for the star-forming central galaxies, although they remain consistent with the field value. The star-forming satellite galaxies have a higher $\mu_{SFR}$ than the field in HB11, which then drops to be closer to the field value in HB12 and HB13. Once again, all the satellite $\mu_{SFR}$ are within error of the field value. As $\mu_{M_{\star}}$ and $\mu_{SFR}$ are larger for the star-forming satellite galaxies than other samples, this suggests that for this halo bin, the peak of C migrates up the galaxy MS for this galaxy sample. The quiescent galaxies have $\mu_{SFR}$ that moves to higher SFR as the environmental mass increases from HB11 to HB12, as can be seen in Fig. \ref{fig:mu-y}b. The central galaxies then remain approximately constant into HB13 while the satellite's $\mu_{SFR}$ moves to a lower SFR.

\begin{figure}
	\resizebox{\hsize}{!}{\includegraphics{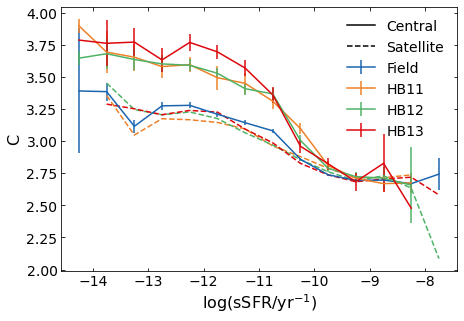}}
	\caption{Median of C as a function of sSFR for the field galaxies (blue) and the central (solid) and satellite (dashed) HB11 (orange), HB12 (green), and HB13 (red) groups. The satellite galaxies closely follow the field galaxies while the central galaxies have higher C at low sSFR. Error bars for satellite galaxies are omitted for clarity of the plot but are comparable to the errors of the central galaxies.}
	\label{fig:roll:C}
\end{figure}

To more clearly see the difference between the central and satellite galaxies, we present the median of C for the field and central and satellite galaxies as a function of specific SFR (SFR/M$_{\star}$, sSFR) in Fig. \ref{fig:roll:C} for the entire galaxy population. As can be seen, at log(sSFR/yr$^{-1}$) below approximately -10, the central galaxies have higher C than the satellite and field galaxies before becoming consistent at higher sSFR. The central HB13 galaxies also appear to have a higher C than the central HB11 and HB12 galaxies. Thus, only the central galaxies' concentration is influenced by environment, with satellite and field galaxies not showing any difference in C. From the M$_{\star}$ and C of our galaxies, we see that there is an increase in C as M$_{\star}$ increases. As M$_{\star}$ is correlated with a galaxy's halo mass, there is an increase in C with the halo mass. Thus, the greater C for central galaxies than the satellite galaxies is likely a result of the central galaxies sitting at, or near, the centre of a larger dark matter halo than the satellite galaxies. At higher sSFR the satellite and central galaxies' C become closer as the number of galaxy-galaxy interactions increases, as will be discussed in Sect. \ref{sec:results:asymmetry}. The general trend of decreasing C with increasing sSFR is also expected as galaxies are known to become more compact as the sSFR decreases \citep{2013ApJ...774...47L, 2018ApJ...853..131L}.

\subsection{Gini}
For Gini, the standard deviation along M$_{\star}$ ($\sigma_{M_{\star}}$) for the star-forming central galaxies rises above that, but consistent within error, of the field galaxies in HB11. This rise is continued in HB12 and HB13, where the $\sigma_{M_{\star}}$ are no-longer within error of the field's value. The star-forming satellite galaxies show a different trend: HB11 is larger than the field value, while HB12 and HB13 are lower and consistent with the field value, as shown in Fig. \ref{fig:a} (dark blue, upward pointing triangles). The quiescent central and satellite galaxies more closely follow each other. Both the central and satellite quiescent galaxies have slightly higher $\sigma_{M_{\star}}$ than the field galaxies in HB11, but with the central galaxies remaining consistent within error. In HB12, the central and satellite galaxies straddle the field's $\sigma_{M_{\star}}$ value, with the central below and satellite above, but are consistent with each other. In the highest halo mass bin, HB13, both the satellite and central galaxies' $\sigma_{M_{\star}}$ are above, but consistent with, the field value.

For the standard deviation along SFR ($\sigma_{SFR}$), the star-forming central galaxies have a slightly higher value than the field galaxies in HB11. This then decreases as the halo mass increases in HB12 before rising again in HB13. The satellite star-forming galaxies' $\sigma_{SFR}$ is also higher than the field galaxies' in HB11 but here it rises into HB12 before falling again in HB13. For the central quiescent galaxies, there is an increase in $\sigma_{SFR}$ compared to the field $\sigma_{SFR}$ in HB11. It rises further into HB12 before $\sigma_{SFR}$ falls in HB13, but all three environments are consistent with one another. For the quiescent satellite galaxies, $\sigma_{SFR}$ is lower in HB11 than the field and rises through HB12 into HB13.

The star-forming galaxies' Gini are influenced by the environmental mass, although this influence is non-trivial for how Gini changes with M$_{\star}$. For the change of Gini with SFR, the central again have a non-trivial relation while the satellite galaxies indicate a slight increase in $\sigma_{SFR}$ as the halo mass increases resulting in a slower change of Gini as SFR changes. For the quiescent galaxies, the rate of change of Gini with M$_{\star}$ is again difficult to find a precise trend, with an approximately constant $\sigma_{M_{\star}}$ for the central galaxies, or constant variability with M$_{\star}$ as the halo mass increases. The satellite galaxies show a very weak decrease in $\sigma_{M_{\star}}$ as the halo mass increases, indicating that the variability along M$_{\star}$ may increase slightly with halo mass. Along the SFR, the satellite galaxies show decreasing variability along SFR as the halo mass increases while the central galaxies show approximately constant variability of Gini along SFR.

The correlation between SFR and M$_{\star}$ for the star-forming galaxies ($\rho$) remains consistent across all environments for both the central and satellite galaxies. The exception is for the star-forming satellite galaxies in HB13, which has a much lower $\rho$ than the field and other environment. For the quiescent galaxies, there is a weak indication that $\rho$ increases for central galaxies as the halo mass increases, which is not seen for the quiescent galaxies.

The absolute values of $A_{G}$ for both the star-forming central and satellite galaxies are consistent with the absolute $A_{G}$ of the field galaxies and so the ranges of Gini are not influenced by the environment. This is also true for the true values of $A_{G}$, with all $A_{G}$ being negative. As $A_{G}$ is negative, $A_{0}$ describes the maximum Gini coefficient while $A_{0}+A_{G}$ describes the minimum Gini. The minimum Gini is consistent across all environments for the central and satellite galaxies. This is also seen for the maximum values of Gini with all halo environments being within error of the field. However, the values of $A_{0}$ are slightly lower in the group environments than the field for both central and satellite galaxies. This may hint that the most compact galaxies are slightly less compact in group environments.

For the quiescent galaxies, the absolute values of $A_{G}$, and indeed the true values, for the central galaxies are consistent across the halo environments, but are typically larger than that of the field galaxies, implying there is a larger range of Gini in the halo environments for central galaxies. For the quiescent satellite galaxies, the absolute value of $A_{G}$ is again consistent across the different halo masses and are also consistent with the field galaxies. There is an indication that the central galaxies have a wider range of Gini than the satellite galaxies. As $A_{G}$ are all negative, $A_{0}+A_{G}$ describes the minimum Gini and $A_{0}$ the maximum in all environments. For the quiescent central galaxies, there is a reduction in the minimum value of Gini in group environments compared to the field. Although, due to large uncertainties all group halo mass bins are within error of the field. The quiescent satellite galaxies, however, do not show the same reduction of the minimum in the group environments. The maximum Gini values for the quiescent central galaxies do not show the reduction seen in the minimum value when compared to the field, although there is a slight drop in HB12. For the satellite galaxies, HB11 is slightly larger than the field while HB12 and HB13 take similar values to the field. Again, due to large uncertainties, all are within error of the field.

The mean along M$_{\star}$ ($\mu_{M_{\star}}$) for the star-forming satellite galaxies remains consistent in all group environments as well as the field. $\mu_{M_{\star}}$ for the star-forming satellite galaxies moves to lower M$_{\star}$ in HB12 and HB13. For the quiescent galaxies, $\mu_{M_{\star}}$ is lower than the field for the central galaxies while the central galaxies are lower than, but consistent with, the field. The satellite galaxies also show a slight decrease in $\mu_{M_{\star}}$ in HB12. The mean along SFR ($\mu_{SFR}$) for the star-forming satellite galaxies have a weak indication that $\mu_{SFR}$ is migrating to higher SFR as the halo mass increases. This is not seen for the central galaxies, where $\mu_{SFR}$ moves to lower SFR in HB12 before moving to slightly higher SFR in HB13. The quiescent galaxies, both central and satellite, have $\mu_{SFR}$ moving to lower SFR as the halo mass increases, where the central galaxies have $\mu_{SFR}$ at higher SFR than the satellites. Thus, the position of the maximum or minimum Gini (depending on the sign of $A_{G}$) does not migrate is a simple way across the SFR-M$_{\star}$ plane as the environment becomes more massive. The movement for the peak of Gini for the star-forming central galaxies does, however, appear to migrate up or down the MS, with both $\mu_{M_{\star}}$ and $\mu_{SFR}$ becoming larger or smaller together.

\begin{figure}
	\resizebox{\hsize}{!}{\includegraphics{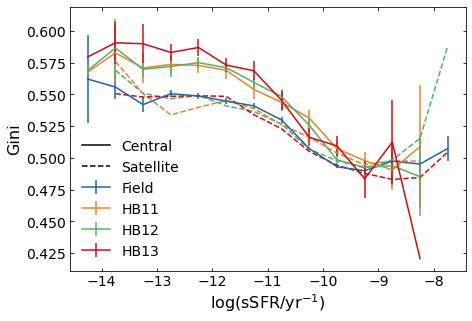}}
	\caption{Median of Gini as a function of sSFR for the field galaxies (blue) and the central (solid) and satellite (dashed) HB11 (orange), HB12 (green), and HB13 (red) groups. The satellite galaxies closely follow the field galaxies while the central galaxies have higher Gini at low sSFR. Error bars for satellite galaxies are omitted for clarity of the plot but are comparable to the errors of the central galaxies.}
	\label{fig:roll:Gini}
\end{figure}

Like C, we plot the median of Gini as a function of sSFR in Fig. \ref{fig:roll:Gini}. As with C, we find that the median Gini is higher for the central galaxies than the satellite galaxies, which are again consistent with the field, bellow approximately log(sSFR/yr$^{-1}$) = -10. Again, we see that this increase for the central galaxies compared to the field is greater for HB13 than the other group environments. Thus, like with C, there is an increase in the compactness of the galaxy's emission for the central galaxies when compared to the satellite and field galaxies. Again, this increase in Gini for low sSFR central galaxies may be due to the central galaxies sitting in, or near, the centre of a large dark matter halo while the satellite galaxies are further away from the centre and lie in smaller sub-halos. The median of Gini also demonstrates how galaxies become more diffuse as the sSFR increases.

\subsection{M$_{20}$}
For M$_{20}$, the $\sigma_{M_{\star}}$ for the star-forming central galaxies are lower than the field in HB11 and HB13, although the field value is within error of HB13. For the star-forming satellite galaxies, all group environments have lower $\sigma_{M_{\star}}$ than the field but due to their large uncertainties, HB11 and HB12 are within error of the field as can be seen in Fig \ref{fig:a}a (light blue downward pointing triangles). For the quiescent galaxies, all halo mass groups are within error of the field value. However, there is an indication that $\sigma_{M_{\star}}$ is decreasing as the group halo mass increases, from above the field in HB11 to approximately equal in HB13.

The standard deviation along SFR shows a marked decrease for the star-forming central and satellite galaxies in HB11. $\sigma_{M_{\star}}$ then rises in HB12, still remaining below the field, and continues to rise in HB13 for the satellite galaxies but falls slightly for the centrals. For the quiescent galaxies, the opposite is seen. In HB11, the $\sigma_{SFR}$ for the central and satellite galaxies are higher than the field and both become smaller in HB12. In HB13, the satellite galaxies' $\sigma_{SFR}$ continues to decrease while that of the centrals increases but remains lower than the value in HB11.

Thus, for the star-forming galaxies, both central and satellite, the change of M$_{20}$ along M$_{\star}$ does not have a simple trend with environment. This is also true for the trend along SFR. For the quiescent galaxies, however, there is an indication that the variability along M$_{\star}$ increases as the group halo becomes larger, seen in the decrease in $\sigma_{M_{\star}}$, while it decreases along SFR. The opposite is seen for the satellite galaxies with decreasing variability along M$_{\star}$ and increasing variability of M$_{20}$ along SFR.

The correlation between SFR and M$_{\star}$ for the star-forming galaxies remains approximately constant and consistent in the field and all group environments for both central and satellites, with the satellite galaxies having slightly lower $\rho$ than centrals. For the quiescent galaxies, $\rho$ is more variable, with the central galaxies' $\rho$ showing a weak reduction as the group halo mass increases, starting higher than the field in HB11. The quiescent satellite galaxies See a reduction in $\rho$ from HB11 to HB12 before it rises slightly in HB13.

The absolute values of $A_{G}$ for the star-forming central HB11 and HB13 are consistent with the field. The central HB12 galaxies have an absolute $A_{G}$ that is lower. For the star-forming satellite galaxies, the absolute $A_{G}$ is lower than the field in HB11 and rises through HB12 to HB13, the last two of which are within error of the field. Thus the range of M$_{20}$ increases with group halo mass for satellite galaxies but not centrals. The absolute values of $A_{G}$ are all positive, for both star-forming and quiescent galaxies. As such, $A_{0}$ describes the minimum value of M$_{20}$ and $A_{0}+A_{G}$ the maximum. For the star-forming galaxies, both the central and satellite's $A_{0}+A_{G}$ remain constant in all environments, including the field, indicating that the most compact galaxies are not further compressed when in groups. The $A_{0}$, therefore, follow similar trends to $A_{G}$, with the central galaxies having a higher value in HB12 corresponding to the slightly lower $A_{G}$ seen for this sample. The satellite galaxies show a decreasing $A_{0}$ as the halo mass increases, becoming consistent with the field in HB13. So while the most compact galaxies are not influenced by environment, the least compact satellite galaxies become less compact as halo mass increases.

For the quiescent galaxies, $A_{G}$ are again all positive and the satellite galaxies show an increase from HB11 to HB12 and a large drop to below the field value in HB13. The quiescent central galaxies show an increase in HB11, compared to the field, but a decrease to below the field in HB12 and HB13. The HB12 and HB13 centrals' $A_{G}$ are approximately the same value. The minimum values of M$_{20}$, here described by $A_{0}$, are consistent for the central galaxies at all halo masses. However, the $A_{0}$ values for the satellite galaxies echo what is seen in $A_{G}$, with a decreasing value from HB11 to HB12 before an increase to above the field value in HB13. Thus, the least compressed galaxies become less compressed as the group halo mass increases. The maximum value of the quiescent satellites remains approximately constant in all environments, group and field. The central galaxies are constant in the field, HB12 and HB13 but have a larger maximum in HB11.

While the $\mu_{M_{\star}}$ of the star-forming galaxies appear to be consistent between the field and all group galaxies, the central HB12 $\mu_{M_{\star}}$ is at a notably lower M$_{\star}$ than the other central group galaxies. The satellite HB11 galaxies are within error of the field and HB12 and HB13 satellites but it is again at a lower M$_{\star}$, as seen in Fig. \ref{fig:mu-x}a. The quiescent galaxies show $\mu_{M_{\star}}$ moving to higher M$_{\star}$ as the group halo mass increases, starting below the field value in HB11 before rising above the field in HB12 for the centrals and HB13 for the satellites. The $\mu_{SFR}$ of the star-forming galaxies show a similar trend to $\mu_{M_{\star}}$ for the central galaxies: $\mu_{SFR}$ moves to a lower SFR from HB11 to HB12 before moving to higher SFR in HB13. For the star-forming satellite galaxies, $\mu_{SFR}$ appears to migrate to higher SFR as the group halo mass increases. The quiescent satellite galaxies do not show this, with $\mu_{SFR}$ remaining approximately constant across all halo mass bins although slightly higher than the field. The quiescent central galaxies also appear to have $\mu_{SFR}$ that is independent of the group mass. Thus it appears that the position of the maximum value of M$_{20}$ on the SFR-M$_{\star}$ plane is approximately independent of the mass of the environment, with only the quiescent galaxies showing a consistent migration to higher M$_{\star}$ as the group mass increases.

\begin{figure}
	\resizebox{\hsize}{!}{\includegraphics{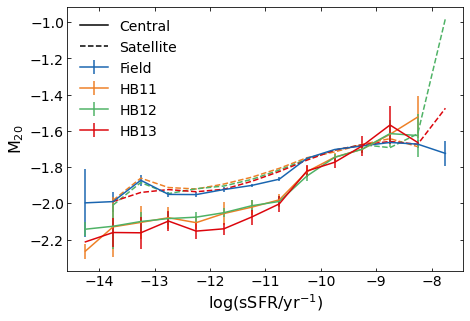}}
	\caption{Median of M$_{20}$ as a function of sSFR for the field galaxies (blue) and the central (solid) and satellite (dashed) HB11 (orange), HB12 (green), and HB13 (red) groups. The satellite galaxies closely follow the field galaxies while the central galaxies have higher M$_{20}$ at low sSFR. Error bars for satellite galaxies are omitted for clarity of the plot but are comparable to the errors of the central galaxies.}
	\label{fig:roll:m20}
\end{figure}

Examining the median M$_{20}$ as sSFR changes in Fig. \ref{fig:roll:m20}, we again see the satellite galaxies closely following the field trend while the central galaxies have lower M$_{20}$ at log(sSFR/yr$^{-1}$) $\lesssim$ -9.5. The difference between group environments is not as clear as with C and Gini, with HB11 and HB13 appearing to agree at log(sSFR/yr$^{-1}$) $\lesssim$ -13.5 before HB11 rises to meet HB12 below log(sSFR/yr$^{-1}$) $\approx$ -12 followed by HB13 rising to meet the other two group bins at log(sSFR/yr$^{-1}$) $\approx$ -11. M$_{20}$ is thus continuing the trend seen with C and Gini. Central galaxies are more compact (have lower M$_{20}$) than the satellite and field galaxies, again likely a result of the central galaxies sitting in the heart of a large dark matter halo. As with C and Gini, the general trend of increasing M$_{20}$ as sSFR increases shows that galaxies with more star-formation activity are less compact.

\subsection{Asymmetry}\label{sec:results:asymmetry}
The $\sigma_{M_{\star}}$ for A are consistent within error for the star-forming satellite galaxies while the star-forming central galaxies are not. The star-forming central galaxies have $\sigma_{M_{\star}}$ that is larger than the field, and satellite galaxies, in all three group halo mass bins. There is a decrease in $\sigma_{M_{\star}}$ from HB11 to HB12 and a large increase in HB13. For the quiescent galaxies, the satellites are again consistent with the field value. However, for the quiescent galaxies the central HB11 and HB13 galaxies are also consistent with the field while HB12 takes a value that is larger.

Along SFR, $\sigma_{SFR}$ in all group environments for the star-forming central and satellite galaxies are consistent, within error, with $\sigma_{SFR}$ of the field galaxies, as can be seen in Fig. \ref{fig:a}a (green stars). There is a weak indication that the $\sigma_{SFR}$ for the central galaxies is increasing with group halo mass. For the quiescent galaxies, the central galaxies are again consistent with the field in all halo mass bins, although $\sigma_{SFR}$ is slightly larger in HB13. The satellite galaxies, however, have a higher $\sigma_{SFR}$ than the field in all group environments.

As a result, the change of asymmetry along M$_{\star}$ is likely uninfluenced by the environment for both the star-forming and quiescent satellite galaxies. The star-forming central galaxies are likely to be uninfluenced by their group but have systematically lower variability in A along M$_{\star}$ than the satellite galaxies. There is a weak indication that the variability along SFR decreases with increasing halo mass for the quiescent central galaxies. The satellite galaxies' variability along SFR does not seem to change as the group mass changes although the variability is less than the field for the quiescent galaxies.

For the correlation between SFR and M$_{\star}$, the star-forming central galaxies are consistent with the field in HB11 and HB12 and lower in HB13. The star-forming satellite galaxies show the opposite: lower than the field in HB11 and HB12 but consistent with the field in HB13. For the quiescent galaxies, the satellite HB11 and central HB13 galaxies are consistent with the field's $\rho$. The other galaxy samples have $\rho$ values that are lower than the field, as can be seen in the lower panel of Fig. \ref{fig:b}.

The absolute values of $A_{G}$ for the star-forming satellite galaxies across all environments are all consistent with one another and the absolute $A_{G}$ in the field. For the star-forming central galaxies, HB11 and HB12 have a larger absolute $A_{G}$, indicating that the galaxies in these two samples are less self similar than in the field or the star-forming central HB13 galaxies as $A_{G}$ describes the range of asymmetry within a population. As all the true values for $A_{G}$ are positive, $A_{0}$ describes the minimum asymmetry in a sample and $A_{0}+A_{G}$ describes the maximum value. For the star-forming galaxies, nearly all environments' $A_{0}$ are again consistent with each other and the field. The central galaxies have an exception, this time with HB13 having a lower $A_{0}$ than the other environments. For the maximum values of asymmetry, $A_{0}+A_{G}$, both the central and satellite galaxies in all environments are consistent with the field value.

As with the star-forming galaxies, all $A_{G}$ of the quiescent galaxies are positive. However, the $A_{G}$ of the group galaxies are not all consistent with the field, with only the satellite HB11 and HB13 and central HB12 galaxies being consistent with the field. Otherwise, the $A_{G}$ are smaller in the group environments than in the field, showing that the range of asymmetry in group environments is typically smaller than in the field. The $A_{0}$, here describing the minimum asymmetry in a sample, are consistent with the field for all central and HB11 and HB12 satellites, with HB13 satellite galaxies having a lower $A_{0}$. For the maximum asymmetry, $A_{0}+A_{G}$, the central galaxies are all consistent with the field environment while the satellite galaxies see a slight reduction in the maximum asymmetry as the group mass increases.

Due to large uncertainties, the $\mu_{M_{\star}}$ for all star-forming galaxies remain within the error of one another in all environments. There is a drop in HB13 for the satellite galaxies as well as a rise at the same group mass for the centrals. For the quiescent galaxies, all group environments have $\mu_{M_{\star}}$ at lower M$_{\star}$ than the field environment. The central galaxies are approximately constant between HB11 and HB12 before rising in HB13 while the satellites rise from HB11 to HB12 before dropping slightly in HB13. The star-forming galaxies again are all consistent with each other across all environments, group and field, for $\mu_{SFR}$. As with $\mu_{M_{\star}}$, the quiescent galaxies again have $\mu_{SFR}$ at lower SFR in groups than in the field. However, for $\mu_{SFR}$  the quiescent central galaxies are consistent with the field in HB11 and HB12 before $\mu_{SFR}$ moves to lower SFR in the highest mass groups. The satellite galaxies' $\mu_{SFR}$ moves similarly to their $\mu_{M_{\star}}$: rising between HB11 and HB12 before falling in HB13 as can be seen in Fig. \ref{fig:mu-y}b. Thus the position of the peak of asymmetry appears to remain in the same position for the star-forming galaxies. The position of the maximum migrates around the SFR-M$_{\star}$ plane for the quiescent galaxies but not in a way that appears to be group mass dependant.

Asymmetry has a median that agrees in all environments for low sSFR, unlike the parameters discussed so far. At log(sSFR/yr$^{-1}$) $\approx$ -11.5, the central galaxies see an increase in A while the satellites again follow the field and remain lower. This is possibly a result of the higher SFR of these central galaxies being a result of galaxy-galaxy interactions, which would act to increase the asymmetry of the galaxies involved. It would be expected that an increase in interactions for centrals would also require an increase in interactions for satellite galaxies and the associated rise in asymmetry and SFR. There is an indication of a slight rise in A for the high sSFR satellite galaxies and it may be possible that the minority of satellite galaxies are interacting resulting in this small change as only a minority of the population experiences the increase to A.

\begin{figure}
	\resizebox{\hsize}{!}{\includegraphics{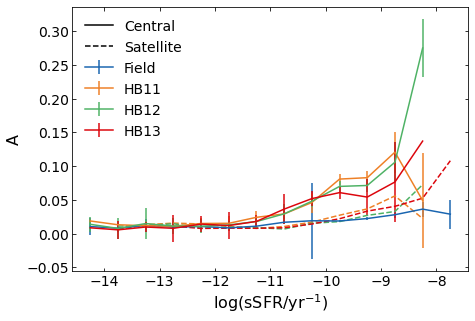}}
	\caption{Median of A as a function of sSFR for the field galaxies (blue) and the central (solid) and satellite (dashed) HB11 (orange), HB12 (green), and HB13 (red) groups. The satellite galaxies closely follow the field galaxies while the central galaxies have higher A at high sSFR. Error bars for satellite galaxies are omitted for clarity of the plot but are comparable to the errors of the central galaxies.}
	\label{fig:roll:asymmetry}
\end{figure}

\subsection{Gini-M$_{20}$ bulge}
For the Gini-M$_{20}$ bulge, $\sigma_{M_{\star}}$ appears to decrease with increasing group halo mass for the star-forming central galaxies, starting higher than the field in HB11 and becoming approximately the same as the field in HB13. This is not seen in the satellite galaxies, that appear to remain constant in all environments and are consistent with the field. This is not seen for the quiescent galaxies where the $\sigma_{M_{\star}}$ for the quiescent galaxies remain consistent and approximately constant in all environments for the central and satellite galaxies as well as the field.

Along SFR, the $\sigma_{SFR}$ of the star-forming central galaxies appears to follow what is seen along M$_{\star}$, $\sigma_{SFR}$ reducing as the group halo mas increases. Although here the larger uncertainties make this trend weak. The star-forming satellite galaxies show no dependence on environment, again like the results from $\sigma_{M_{\star}}$. The quiescent galaxies are all within error of the field for both the central and satellite galaxies. However, both the central and satellite galaxies are lower than the field in HB11, rise to be higher in HB12, and then drop in HB13.

This implies that the variability of GMB along M$_{\star}$ and SFR for the star-forming central galaxies increases with halo mass increases. The quiescent central galaxies do not see this, with the variability remaining approximately group mass independent. For the star-forming and quiescent satellite galaxies, there is no change in $\sigma_{M_{\star}}$ or $\sigma_{SFR}$ as the halo mass changes.

The correlation between SFR and M$_{\star}$ remains approximately constant for all environments and for central and satellite galaxies. The star-forming galaxies do appear to have a slightly higher $\rho$ than the field for both the central and satellite galaxies, but all are consistent with the field as can be seen in Fig. \ref{fig:b}b (red diamonds).

The absolute $A_{G}$ values for the star-forming group galaxies are all consistent with one another across the three halo mass bins, showing that the range of GMB does not change for star-forming group galaxies. Only the star-forming central galaxies in HB11 are not within error of the field value. The field's true value is negative along with the central HB11 and HB13 samples, unlike the positive $A_{G}$ for all the group remaining environments, as can be seen in Fig. \ref{fig:AG}a. The minimum GMB values for all the star-forming galaxies are all consistent with one another in all group environments, with the field being slightly lower than the groups. The maximum values of GMB for the central and satellite galaxies are all lower than the maximum value for the field, although the values' uncertainties are such that all but the central HB11 are consistent with the field value.

For the quiescent galaxies, the absolute $A_{G}$ values of the satellite galaxies increase with increasing halo mass but remain smaller than the absolute $A_{G}$ of the field, showing that the range of GMB is smaller in group environments than the field. The absolute $A_{G}$ of the central galaxies are also smaller than the field but are consistent. The satellite HB11 is the only sample whose $A_{G}$ is not positive, meaning this is the only sample where $A_{0}$ is the largest GMB and $A_{0}+A_{G}$ is the smallest. For the minimum values of GMB for the quiescent galaxies, both the central and satellite galaxies in all group mass bins are negative and consistent with the minimum GMB in the field. For the maximum GMB, the central galaxies are consistent with the field value and have weak evidence of an increase with group halo mass. For the satellite galaxies, the maximum value increases with halo mass but remains lower than the field value.

The $\mu_{M_{\star}}$ for the star-forming satellite galaxies are at higher M$_{\star}$ than the field and appear to show a decrease in M$_{\star}$ as the group mass increases. However, due to large uncertainties all three halo mass bins are consistent with one another. The star-forming central galaxies in the lowest and highest group halo mass bins are consistent with the field's $\mu_{M_{\star}}$ while HB12 has a higher value. For the quiescent galaxies, $\mu_{M_{\star}}$ moves to slightly lower M$_{\star}$ from HB12 to HB13 for both the central and satellite galaxies, although the values are within error. In the lower group mass bin, both the central and satellite galaxies are much lower than the field or higher mass bins' $\mu_{M_{\star}}$ but the central galaxies' large uncertainty make it within error of the field and higher mass bins. Along SFR, the star-forming central and satellite galaxies' $\mu_{SFR}$ are all consistent with the field at all group masses. The quiescent galaxies have $\mu_{SFR}$ that is lower than the field's $\mu_{SFR}$, rising between HB11 and HB12 before falling in HB13 as can be seen in Fig. \ref{fig:mu-y}b. Thus, much like asymmetry, the position of the peak of GMB appears to remain in the same position for the star-forming galaxies. For the higher mass groups, the position of the peak does not appear to move much between the group mass bins but is at lower SFR and M$_{\star}$ for the lowest group masses of HB11.

\begin{figure}
	\resizebox{\hsize}{!}{\includegraphics{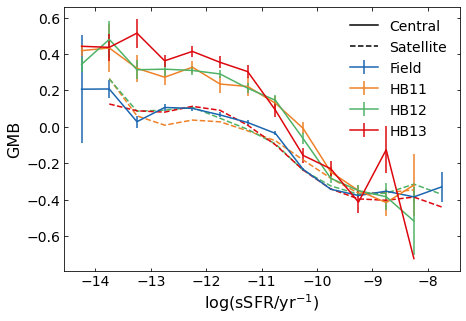}}
	\caption{Median of GMB as a function of sSFR for the field galaxies (blue) and the central (solid) and satellite (dashed) HB11 (orange), HB12 (green), and HB13 (red) groups. The satellite galaxies closely follow the field galaxies while the central galaxies have higher GMB at low sSFR. Error bars for satellite galaxies are omitted for clarity of the plot but are comparable to the errors of the central galaxies.}
	\label{fig:roll:GMB}
\end{figure}

The median of GMB sees a decrease as the sSFR increase for all galaxies, as seen in Fig. \ref{fig:roll:GMB}. Once again, we see that the satellite galaxies closely follow the field galaxies while the central galaxies start with a higher GMB at low sSFR. At the lowest sSFR, it appears that the central HB13 galaxies have higher GMBs than the central galaxies in the other group environments. By log(sSFR/yr$^{-1}$) $\approx$ -9, the central galaxies' GMB has fallen to meet the satellite and field galaxies. The general trends of GMB are what are classically expected, with higher sSFR galaxies showing a more disk-like GMB and lower sSFR more bulge-like. The lower sSFR central galaxies are more bulge-like than the satellite and field lower sSFR galaxies.

\subsection{$r_{50}$}
For star-forming galaxies, the standard deviation of the radius containing 50\% of the light along M$_{\star}$ decreases as the host halo mass increases for both central and satellite galaxies, as can be seen in Fig. \ref{fig:a}a (yellow crosses), remaining much lower than the $\sigma_{M_{\star}}$ of the field at all halo masses. For the quiescent galaxies, $\sigma_{M_{\star}}$ again decreases as the halo mass increases for both centrals and satellites. Here, the field's $\sigma_{M_{\star}}$ is lower than both the central and satellite HB11.

The standard deviation along SFR, $\sigma_{SFR}$, in group environments is much lower than in the field, for the star-forming galaxies. The central star-forming galaxies have $\sigma_{SFR}$ that decreases between HB11 and HB12 and increases slightly between HB12 and HB13 while the satellite galaxies show a very slight decrease in $\sigma_{SFR}$ as the group mass increases. For the quiescent galaxies, $\sigma_{SFR}$ falls as the halo mass increases for the central galaxies, starting higher than the field in HB11. For the quiescent satellite galaxies, there is a rise between HB11 and HB12, the former of which is consistent with the field, before a sharp drop in HB13.

Thus, the variability of $r_{50}$ increases along M$_{\star}$ for all star-forming galaxies while the star-forming central galaxies have group mass independent variability along SFR. The change of variability for the quiescent galaxies increases as halo mass increases along M$_{\star}$, while along SFR it is less clear with the $\sigma$ values increasing or decreasing between halo mass bins for the satellite galaxies.

The $\rho$ for the star-forming galaxies is higher than the field value for both the central and satellite galaxies. The general trend for the central galaxies is an increase in $\rho$ with group mass while the satellite galaxies' $\rho$ is lower in HB12 than HB11 and HB13. For the quiescent galaxies, $\rho$ for the central galaxies again increases with group mass while the satellites' $\rho$ decreases.

The $A_{G}$ for the star-forming central galaxies increases as the halo mass increases, remaining lower than the field, indicating a corresponding increase in the range of $r_{50}$. The $A_{G}$ of the satellite galaxies remains below that of the field and centrals in all halo mass bins, reducing between HB11 and HB12 before increasing again into HB13. The minimum value of $r_{50}$, here described by $A_{0}$, for both the central and satellite galaxies rise between HB11 and HB12 before falling slightly moving from HB12 to HB13 for the central galaxies but continuing to rise for the satellites. The satellite galaxies $A_{0}$ are smaller than the central galaxies, as seen in Fig. \ref{fig:A0}a. The maximum $r_{50}$, described by $A_{0}+A_{G}$, increases for the central galaxies as the halo mass increases, following a similar trend to $A_{G}$ for the central galaxies. The maximum $r_{50}$ for the satellite galaxies are smaller than the field at all group halo masses. There is a slight rise moving from HB11 to HB12 before $A_{0}+A_{G}$ rises significantly in HB13.

For the quiescent galaxies, the $A_{G}$ for the group galaxies rise with group halo mass, more slowly for the satellite galaxies. This implies that the range of $r_{50}$ in groups increases as the group mass increases for all quiescent galaxies. The central galaxies in HB13 are the only sample whose $A_{G}$ rises above the field value. As with the star-forming galaxies, $A_{0}$ describes the minimum $r_{50}$ for the quiescent galaxies and $A_{0}+A_{G}$ the maximum. The minimum of the quiescent central galaxies again rises with group halo mass, from below the field value in HB11 to above it in HB12 and HB13. The satellite galaxies' minimum follows a similar trend. The maximum for all group galaxies, except the central HB13, are below the maximum $r_{50}$ in the field. As with the minimum, the maximum values increase with increasing halo mass for both the central and satellite galaxies, indicating that the largest galaxies become larger as the halo mass increases.

The position of $\mu_{M_{\star}}$ for the star-forming galaxies are all consistent with the value in the field and appear to remain constant in all environments. This is also seen for the quiescent galaxies. For $\mu_{SFR}$, the value in the star-forming field galaxies is at a higher SFR than for the star-forming group galaxies, both centrals and satellites. In the groups for both the central and satellite galaxies, $\mu_{SFR}$ reduces between HB11 and HB12 before rising again in HB13. This is not seen for the quiescent galaxies, where the satellite galaxies in HB11 and HB12 are consistent with the field value before dropping to lower SFR in HB13. The central galaxies have $\mu_{SFR}$ at lower SFR than the field in all group environments. For the quiescent centrals, $\mu_{SFR}$ evolves similarly to the star-forming galaxies: falling from HB11 to HB12 and rising again into HB13. Thus the position of the maximum $r_{50}$ only moves along the SFR axis in the SFR-M$_{\star}$ plane. The maximum of all central galaxies is at lower SFR at intermediate group masses, along with the star-forming satellite galaxies. For the quiescent satellites, the maximum $r_{50}$ only moves to lower SFR for the highest mass group galaxies.

\begin{figure}
	\resizebox{\hsize}{!}{\includegraphics{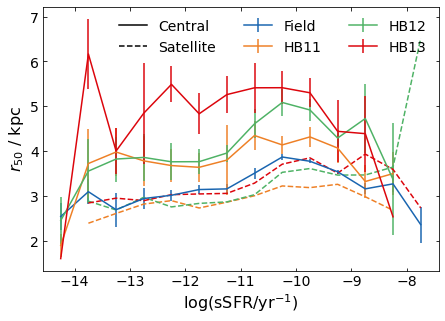}}
	\caption{Median of $r_{50}$ as a function of sSFR for the field galaxies (blue) and the central (solid) and satellite (dashed) HB11 (orange), HB12 (green), and HB13 (red) groups. The satellite galaxies closely follow, or are slightly below, the field galaxies while the central galaxies have higher $r_{50}$ that indicates an increase in $r_{50}$ as the halo mass increases. Error bars for satellite galaxies are omitted for clarity of the plot but are comparable to the errors of the central galaxies.}
	\label{fig:roll:r50}
\end{figure}

The median $r_{50}$ as a function of sSFR is the only parameter where the satellite and field galaxies do not become consistent at some point, with the exception of the highest and lowest sSFR that are likely dominated by low number statistics. The satellite galaxies of HB13 and HB12 closely follow the trend of the field galaxies, as can be seen in Fig. \ref{fig:roll:r50}. HB11 has a slightly lower $r_{50}$ than the other environments at log(sSFR/yr$^{-1}$) $>$ -11. For the central galaxies, as the environment becomes more massive, the median $\mathbf{r_{50}}$ of the galaxies increases, although again HB11 and HB12 agree at log(sSFR/yr$^{-1}$) $<$ -11. The increase in $r_{50}$ of the central galaxies as the halo mass increases is not surprising. More massive group halos are likely to have a more massive central galaxy and the associated larger physical size. Satellite galaxies are likely to be smaller than the central galaxy and their mass is less dependant on the group halo mass, hence why the median of the satellite galaxies are lower than the centrals and more in line with the field galaxies. There is also a slight increase in $r_{50}$ for the field, HB11, and HB12 galaxies, both central and satellite, as sSFR increases, which supports the idea that galaxies with higher sSFRs are more diffuse.

\subsection{Total S\'{e}rsic index}
We find that the standard deviation of $n$ along M$_{\star}$ for the star-forming central galaxies decreases as the environment mass increases, starting above the $\sigma_{M_{\star}}$ for the field in HB11 before becoming consistent with the field in HB13. The $\sigma_{M_{\star}}$ for the satellite galaxies, on the other hand, starts consistent with the field in HB11, rises in HB12 before reducing in HB13, as can be seen in Fig. \ref{fig:a}a (orange squares). For the quiescent galaxies, both the central and satellite galaxies have $\sigma_{M_{\star}}$ higher than the field in HB11 and lower than the field in HB12, with the satellite galaxies having a higher $\sigma_{M_{\star}}$ in both halo mass bins. In HB13, the galaxy types split, with the centrals having $\sigma_{M_{\star}}$ greater than the field and the satellites less than the field.

Like $\sigma_{M_{\star}}$, $\sigma_{SFR}$ of the star-forming central galaxies decreases as the group halo mass increases and becomes consistent with the field in HB13. For the star-forming satellite galaxies, $\sigma_{SFR}$ is also consistent with the field in HB11 before rising in HB12 and HB13. For the quiescent galaxies, both the central and satellite galaxies are constant in HB11 and HB12, with the satellite galaxies having a smaller $\sigma_{SFR}$, before rising slightly in HB13 for the satellites and above the field for the centrals. In all other halo bins, the $\sigma_{SFR}$ of the group galaxies are smaller than the $\sigma_{SFR}$ of the field galaxies.

This results in the variability of $n$ as M$_{\star}$ changes increasing for both the central and satellite star-forming galaxies. This is also seen for $n$ along SFR for the star-forming central galaxies while the star-forming satellites become less variable as the group halo mass increases. The variability of $n$ along SFR appears independent of the halo mass for all quiescent galaxies. Trends along M$_{\star}$ for the quiescent galaxies are less well defined.

For the correlation between SFR and M$_{\star}$, the $\rho$ for the star-forming central galaxies is higher in HB11 and HB12 than the field, before dropping below the field in HB13. For the star-forming satellite galaxies, $\rho$ in HB11 is consistent with the field while HB12 and HB13 are larger. For the quiescent galaxies, the central galaxies show a large drop in HB12 while the satellites show a similar drop in HB11.

For the star-forming galaxies, the absolute value of $A_{G}$ increases for the central galaxies as the halo mass increases, starting larger than the absolute value of $A_{G}$ in the field environment. For the star-forming satellite galaxies the opposite is seen, with the absolute $A_{G}$ decreasing with halo mass from being consistent with the field environment in HB11 to smaller than the field in HB13. Thus the range of $n$ for star-forming central galaxies increases while the range for satellite galaxies decreases. As the true values of $A_{G}$ are all negative, $A_{0}$ is the maximum value for $n$ and $A_{0}+A_{G}$ is the minimum value. For the star-forming central galaxies, the maximum $n$ increases with group halo mass, from slightly higher than the field in HB11 and HB12 followed by a larger rise in HB13. The minimum $n$ for the star-forming central galaxies remains approximately constant across all environments. The maximum $n$ for the star-forming satellite galaxies appears to decrease as the halo mass increases, moving from just above the field value in HB11 to below it in HB13, as seen in Fig. \ref{fig:A0}a. The minimum value of the star-forming satellite galaxies has a slight rise in HB12 but otherwise remains approximately constant.

The absolute value of $A_{G}$ for the quiescent galaxies is high in HB11 when compared to the value in the field environment, for both the central and satellite galaxies. It then drops to become consistent with the field environment in HB12, for the satellites, and HB13, for both the centrals and satellites. Thus, only the range of $n$ in HB11 is larger than the field. As with the star-forming galaxies, all $A_{G}$ values are negative resulting in $A_{0}$ describing the maximum $n$ and $A_{0}+A_{G}$ describing the minimum $n$. For the quiescent central galaxies, the maximum $n$ rises from HB11 to HB12 before falling in HB13. For all groups, $A_{0}$ is larger than the value in the field. The minimum value for these galaxies is lower than the field in HB11, higher than the field in HB12 and then only slightly larger than the field in HB13. For the quiescent satellite galaxies, the minimum $n$ remains consistent with the field in HB11 and HB12 before rising in HB13. The maximum $n$ of the quiescent satellite galaxies is also consistent with the field in HB12, slightly higher than the field in HB13 and much larger in HB11.

The mean along M$_{\star}$ is less than the field galaxies' value in HB11 and HB12 for the star-forming central galaxies and is consistent between these two group mass bins. $\mu_{M_{\star}}$ then rises above the field value in HB13. The star-forming satellite galaxies, however, are consistent with the field in HB11 and HB13 but lower than the field, and consistent with the central galaxies, in HB12 as can be seen in Fig. \ref{fig:mu-x}a. For the quiescent galaxies, only satellite HB12 and central HB13 are consistent with the field $\mu_{M_{\star}}$ while HB11 have $\mu_{M_{\star}}$ at lower M$_{\star}$ and the satellite HB13 at higher M$_{\star}$ than the field. Central galaxies in HB11 are lower than the field and consistent with the satellite HB11 galaxies. The changes with environment for $\mu_{SFR}$ for the star-forming galaxies closely follow the trends seen in $\mu_{M_{\star}}$, with the central galaxies having a lower $\mu_{SFR}$ than the field followed by a rise to HB13. For the satellite galaxies, $\mu_{SFR}$ is also seen to be lower in HB12, as was seen with $\mu_{M_{\star}}$. Unlike $\mu_{M_{\star}}$, the central galaxies see a rise in $\mu_{SFR}$ from HB11 to HB12, starting consistent with the field in HB11, and a drop to become consistent with the field again in HB13. Satellite galaxies have $\mu_{SFR}$ at higher SFRs in all group environments than the field, with HB12 being at a slightly lower SFR than HB11 and HB13.

As a result, the position of the minimum $n$ for the central star-forming galaxies moves to higher M$_{\star}$ and SFR as the group mass increases, from a position with lower M$_{\star}$ and SFR than the field to a position with higher M$_{\star}$ and SFR. The position of the minimums $n$ for the remaining galaxies do not move in such a defined way and appear to wander across the SFR-M$_{\star}$ plane.

\begin{figure}
	\resizebox{\hsize}{!}{\includegraphics{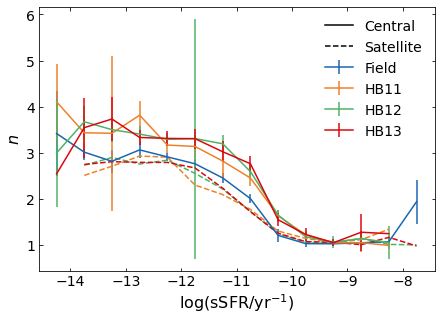}}
	\caption{Median of $n$ as a function of sSFR for the field galaxies (blue) and the central (solid) and satellite (dashed) HB11 (orange), HB12 (green), and HB13 (red) groups. The satellite galaxies closely follow the field galaxies while the central galaxies have higher $n$ at low sSFR. Error bars for satellite galaxies are omitted for clarity of the plot but are comparable to the errors of the central galaxies.}
	\label{fig:roll:n}
\end{figure}

For $n$, the median as a function of sSFR mimics what is seen in GMB, as would be expected as these two parameters describe the disk or bulge domination of a galaxy. The central galaxies' $n$ is higher at low sSFR, as can be seen in Fig. \ref{fig:roll:n}, while the satellites' $n$ closely follows the field. We note, however, that the large uncertainties on the median at lower sSFR mean that this split between the central and satellite galaxies is not strong. All medians drop as sSFR increases and become consistent for centrals and satellites in all environments and the field at log(sSFR/yr$^{-1}$) $\approx$ -9.5. As with GMB, $n$ shows greater bulge domination for the lower sSFR galaxies and greater disk domination for the high sSFR galaxies, as would be expected.

\subsection{Bulge S\'{e}rsic index}
For the final parameter explored in this work, $\sigma_{M_{\star}}$ of the star-forming central galaxies increases with group mass, starting below the field value in HB11 and rising above it in HB13, as can be seen in Fig. \ref{fig:a} (brown pentagrams). The star-forming satellite galaxies have $\sigma_{M_{\star}}$ below the field value in HB11, which increases to be above the field value in HB12. However, for the satellite galaxies, $\sigma_{M_{\star}}$ then reduces to be below the field value in HB13. For the quiescent central galaxies, there is a weak indication that $\sigma_{M_{\star}}$ increases with M$_{star}$ with only the HB12 value not being consistent with the field. For the satellite galaxies, the $\sigma_{M_{\star}}$ is notably larger than any other environment, while HB12 is consistent with the field and HB13 has a $\sigma_{M_{\star}}$ slightly larger than the field.

Along SFR, the $\sigma_{SFR}$ of the star-forming central galaxies is larger than other environments in HB11 while being highly uncertain. In HB12, $\sigma_{SFR}$ is lower than the field value while it is larger than the field in HB13. For the satellite galaxies, the values of $\sigma_{SFR}$ are more certain. $\sigma_{SFR}$ starts consistent with the field in HB11 before falling through HB12 into HB13. The quiescent central galaxies' $\sigma_{SFR}$ is consistent with the field in all group environments, although there is a weak indication that $\sigma_{SFR}$ increases with group mass. $\sigma_{SFR}$ of the HB12 quiescent satellite galaxies is consistent with the field and HB13's $\sigma_{SFR}$ is larger than the field. In HB11, $\sigma_{SFR}$ is notably larger than the field value much like this sample's $\sigma_{M_{\star}}$.

This results in the variability of $n_{Bulge}$ as M$_{\star}$ changes to decrease as the environment becomes more massive for central galaxies, with no clear trends for the satellite galaxies. This is also weakly true for the variability of $n_{Bulge}$ as SFR changes for the quiescent central galaxies while the opposite is true for the star-forming satellites. For the remaining samples, there is again no clear trend for the variability of $n_{Bulge}$ as the SFR changes.

The $\rho$ of the star-forming central galaxies is consistent with the field in HB11 and HB13 while HB12 is larger than the field's value. For the star-forming satellites, $\rho$ is lower than the field's value in all group environments but rises from HB11 to HB12 before becoming approximately constant between HB12 and HB13. For the quiescent central galaxies, there is a weak indication that $\rho$ is increasing with group mass but all three group environments have consistent $\rho$ that are lower than the field value. The quiescent satellite galaxies, on the other hand, show a clear reduction of $\rho$ and the group mass increases, staring above the field value in HB11, becoming consistent with the field in HB12 and falling below the field in HB13.

For the star-forming galaxies, the absolute values of $A_{G}$ increase, for the central galaxies, from HB11 to HB12 before decreasing in HB13. However, due to the large uncertainties seen in Fig. \ref{fig:AG}, both HB11 and HB13 are consistent with each other and the field. For the satellite galaxies, the absolute values of $A_{G}$ in the group environments are larger than the field, with HB12 being the smallest, followed by HB11 then HB13. As all values of $A_{G}$ are negative for the star-forming galaxies, $A_{0}$ describes the maximum $n_{Bulge}$ and $A_{0}$+$A_{G}$ the minimum. The maximum value of $n_{Bulge}$ for the star-forming central galaxies is larger in the field than in all group environments. Within the group environments, the maximum values rises slightly between HB11 and HB12 before remaining constant in HB13. For the minimum of the star-forming central galaxies, the field again has the largest value. HB11 is the next largest, with HB13 being slightly smaller but consistent with HB11. HB12, therefore, has the smallest minimum. The star-forming satellite galaxies have a maximum that is consistent between the field and HB11, before rising into HB12 and falling below the field value in HB13. The minimum $n_{Bulge}$ of the group environments follows a similar trend - rising from HB11 to HB12 before falling in HB13 - but here all environments have a smaller minimum $n_{Bulge}$ than the field.

For the quiescent galaxies, the absolute values of $A_{G}$ decrease with environment mass for the satellite galaxies. The central galaxies have an absolute $A_{G}$ in HB11 that is lower than, but consistent with, the field while the values in HB12 and HB13 are larger than the field and consistent with one another. As with the star-forming galaxies, all $A_{G}$ are negative so $A_{0}$ describes the maximum value and $A_{0}$+$A_{G}$ the minimum $n_{Bulge}$. For the central galaxies, the maximum $n_{Bulge}$ rise as the group mass rises, being consistent with the field in HB12. The minimum rises from HB11 to HB12 where it remains constant into HB13. For the quiescent satellite galaxies, the maximum value is constant and above the field value in HB11 and HB12, before falling below the field value in HB13. The minimum value of $n_{Bulge}$ for the quiescent satellite galaxies falls as the environment becomes more massive, from the field to HB13.

The mean along M$_{\star}$, $\mu_{M_{\star}}$, for the star-forming central galaxies moves to a lower value from HB11 to HB12, the latter of which is consistent with the field, before notably increasing in HB13. For the star-forming satellites, $\mu_{M_{\star}}$ moves to lower values as the environment mass increases, from the field to HB13. $\mu_{M_{\star}}$ for the quiescent central galaxies is at much larger values than the field for all group galaxies, with HB13 being at a slightly lower value than HB11 and HB12, the last two of which are consistent with one another. The quiescent satellites have $\mu_{M_{\star}}$ at a smaller value than the field in HB11, although it is consistent with the field. HB12 has a $\mu_{M_{\star}}$ that is consistent with the field while HB13's value is much larger. For the mean along SFR, $\mu_{SFR}$, the star-forming central galaxies have a notably low value in HB11. HB12 has a $\mu_{SFR}$ value that is lower than the field and $\mu_{SFR}$ decreases further in HB13, with both HB12 and HB13 having values much larger than HB11. For the star-forming satellite galaxies, all $\mu_{SFR}$ values are larger than the field but remain approximately constant across all group environments. For the quiescent central galaxies, $\mu_{SFR}$ is consistent with the field in HB13. In HB11 and HB12 it is lower than the field, although the HB11 value is again consistent with the field. For the quiescent satellite galaxies, $\mu_{SFR}$ starts at a lower value than the field in HB11, rises slightly in HB12, before falling notably in HB13. Thus, the position of the maximum $n_{Bulge}$ does not appear to move in a well defined way across the M$_{\star}$-SFR plane for any of the four galaxy samples.

The median $n_{Bulge}$ with sSFR weakly follows what is seen in GMB and $n$ with the median moving to lower $n_{Bulge}$ as sSFR increases. However, there is no splitting between the central and satellite group populations or the group population and the field, as seen in GMB and $n$. The downward trend is also subject to the large uncertainties that mean this trend is weak. If the trend is truly present, then the lower sSFR galaxies have stronger bulges than the lower sSFR galaxies, again mimicking what has been see with GMB and $n$.

\begin{figure}
	\resizebox{\hsize}{!}{\includegraphics{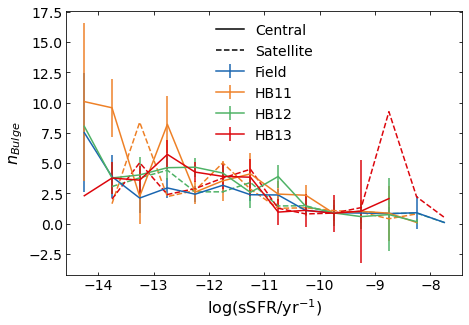}}
	\caption{Median of $n_Bulge$ as a function of sSFR for the field galaxies (blue) and the central (solid) and satellite (dashed) HB11 (orange), HB12 (green), and HB13 (red) groups. The central and satellite galaxies in all environments closely follow the field galaxies. Where the central HB11 galaxies appear to rise above the other environments, the uncertainties are such that this rise is not significant. Error bars for satellite galaxies are omitted for clarity of the plot but are comparable to the errors of the central galaxies.}
	\label{fig:roll:n-bulge}
\end{figure}

\section{Discussion}\label{sec:disc}
\subsection{Compactness of galaxies}
A number of the parameters in this work can be considered to be descriptions of how compact a galaxy is, namely C, Gini, M$_{20}$, $n$ and to a lesser extent $r_{50}$. Thus, it is possible to consider these parameters together to see if they offer a consistent picture.

Looking at the change to the standard deviations of these five parameters, no galaxy group is consistent across all five, even when making the assumption that trends that rise then fall (or fall then rise) as the environment becomes more massive are an indication that there is not an environmental dependence. The star-forming satellite galaxies agree in four parameters, C, Gini, M$_{20}$, and $n$, for $\sigma_{M_{\star}}$, which all show independence from the group environment. The same four parameters also show environmental independence for $\sigma_{SFR}$ for the quiescent central galaxies. A different four parameters, C, Gini, M$_{20}$ and $r_{50}$, also show no relation between group environment and $\sigma_{SFR}$ for the star-forming central galaxies. For the quiescent satellite galaxies, C, M$_{20}$ and $r_{50}$ show $\sigma_{M_{\star}}$ decreases as the group becomes more massive, if the mind decrease in $\sigma_{M_{\star}}$ noted for C is truly there, while C, $r_{50}$ and $n$ show no dependence of $\sigma_{SFR}$ on group mass.

The range of these five parameters again does not show a consistent picture. C, Gini and M$_{20}$ agree that  there is no environmental dependence on the range for the star-forming central and all quiescent galaxies, meaning that the most and least compact galaxies do not become more or less separated. $r_{50}$, on the other hand, shows an increase in the range for all three galaxy samples. The total S\'{e}rsic index shows an increase for the star-forming central galaxies but a decrease for the star-forming satellites and quiescent central galaxies. For the star-forming satellite galaxies, Gini and $r_{50}$ both see no dependence for the range of parameter with group mass while C and M$_{20}$ show an increasing range of parameter with group halo mass and $n$ a decrease. Thus, it appears that the most and least compact galaxies do not change their separation of the majority of galaxies, while the environmental dependence for the star-forming satellite galaxies is unclear.

For the change to the most compact galaxies, it is necessary to compare the maximum values of C, Gini, M$_{20}$, and $n$ with the minimum values of $r_{50}$ as a smaller $r_{50}$ implies a more compact galaxy. Similarly, to study the least compact galaxies we must compare the minimum C, Gini, M$_{20}$ and $n$ and maximum $r_{50}$. Once again, no sample of galaxies is consistent across all five parameters. The star-forming central galaxies show a consistent picture of the most compact galaxies with C, Gini, M$_{20}$ and $r_{50}$, showing no environmental dependence. The most compact C, Gini, M$_{20}$ and $n$ all agree for all quiescent galaxies, again showing no environmental dependence. C and $r_{50}$ show that the most compact star-forming satellite galaxies become less compact as the group mass increases while M$_{20}$ and Gini show no environmental change. For the least compact galaxies, C, Gini, M$_{20}$ and $n$ agree for the star-forming central galaxies and all quiescent galaxies that there is no environment dependence. For the star-forming satellite galaxies, C, M$_{20}$, and $n$ all indicate that there is again no environmental dependence, while Gini indicates the galaxies become less compact and $r_{50}$ indicates the opposite.

Gini and M$_{20}$ are able to describe how clumpy the light distribution is within a galaxy, while C assumes rotational symmetry about a galaxy's centre. As the majority of the galaxies with the highest and lowest Gini or M$_{20}$ do not notably change the value of their parameter for the majority of galaxies, it would appear that the most extreme galaxies do not become clumpier, or less clumpy, as their environment becomes more massive, and potentially disruptive. With the constant maximum and minimum C in all group environments for all but most compact star-forming satellite galaxies, it also appears that the shape of the light profile predominantly remain unchanged.

The lack of agreement from $r_{50}$ indicates that it is a poor descriptor of how compact a galaxy is and instead is describing the size of the galaxy. This does seem to show that the smallest and largest galaxies in groups become larger as the group halo mass increases, for the star-forming satellite and quiescent central and satellite galaxies. The largest star-forming centrals are also becoming larger while the smallest are not.

As C describes the ratio between the radius that contains 20\% of the light ($r_{20}$) and the radius that contains 80\% of the light ($r_{80}$), it can be used with $r_{50}$ to better understand how the galaxies are changing as the group halo mass changes. For the quiescent galaxies and the star-forming central galaxies, the maximum and minimum C remain approximately constant, implying that the ratio between $r_{20}$ and $r_{80}$ remains constant. With $r_{50}$ seen to increase with group halo mass, for both the largest and smallest $r_{50}$ of the quiescent galaxies, this would suggest that the largest and smallest quiescent galaxies themselves are becoming larger. For the star-forming central galaxies, only the smallest $r_{50}$ become larger and so only the smallest galaxies become bigger with group halo mass. This could be explained for the satellite galaxies as the increase in galaxy-galaxy interactions as the halo mass increases causing the galaxies becoming more rarefied and defuse. This would increase $r_{50}$ while not increasing C, or indeed Gini or M$_{20}$ assuming the clumpiness of the galaxies remains unchanged. As the range of $r_{50}$ increases with group halo mass, the largest galaxies become larger faster than the smallest galaxies. For the star-forming satellite galaxies, as C increases for the most compact galaxies along with $r_{50}$, it is possible that the outer regions of the most compact galaxies are expanding but not the inner region. This would result in $r_{50}$ and $r_{80}$ becoming larger but not $r_{50}$. For the least compact galaxies, C is constant and $r_{50}$ increases, so again the smallest galaxies become larger as the halo mass increases.

For the median of the entire galaxy population within each environment - that is both star-forming and quiescent galaxies within the field, HB11, HB12, and HB13 - C, Gini, M$_{20}$, and $n$ show a similar picture. All four of these parameters show that galaxies become more compact as the sSFR decreases in all environments. This increase in compactness is greater for central galaxies than satellite or field galaxies, the last two of which appear to evolve in similar ways for C, Gini, M$_{20}$ and $n$. We reiterate that due to large uncertainties, the split between the central and satellite galaxies seen with $n$ is not strong. For the central galaxies, the HB13 sample appear to be less diffuse than the centrals of HB11 and HB12 for C, Gini, and M$_{20}$. $r_{50}$ again differs in what is seen, showing that is it a poor tracer of the compactness of a galaxy. Central galaxies have larger $r_{50}$ than the field or satellite galaxies with a slight increase as sSFR increases for the field, HB11, and HB12. If $r_{50}$ was describing compactness, this would mean that central galaxies are more diffuse than the field or satellite galaxies at all sSFR, which is not seen for C, Gini, and M$_{20}$. However, the slight rise in $r_{50}$ at higher sSFR for the field, HB11, and HB12 galaxies does support the decrease in compactness as sSFR increases.

\subsection{Disks and bulges}\label{subsec:disk-n-bulge}
The bulge S\'{e}rsic index and Gini-M$_{20}$ bulge both describe if a galaxy is bulge or disk dominated and should show similar trends with environmental mass. If we take the conservative position that trends that undulate, that is the value increases (decreases) from HB11 to HB12 then decreases (increases) from HB12 to HB13, show independence from the group environment, then the $\sigma_{M_{\star}}$ of GMB and $n_{Bulge}$ agree for the satellite galaxies but not the centrals. Both GMB and $n_{Bulge}$ show no dependence of $\sigma_{M_{\star}}$ on group mass for the satellite galaxies. The $\sigma_{M_{\star}}$ of the star-forming central galaxies reduces as the halo mass increases for GMB while the opposite is seen in $n_{Bulge}$. GMB has no dependence on the halo mass for the quiescent central galaxies while $n_{Bulge}$ again shows an increase. Group environments, therefore, predominantly appear to only influence the central galaxies' dependence on M$_{\star}$ but the exact form of this dependence is unclear.

The group mass has no clear influence of the bulge and disk components with respect to the SFR, with only GMB and $n_{Bulge}$ agreeing that there is no dependence for the quiescent satellite galaxies. If the weak trend of $\sigma_{SFR}$ of $n_{Bulge}$ increasing with group mass for the quiescent central galaxies is taken to be so weak as to be constant $\sigma_{SFR}$ with group mass, then the quiescent central galaxies could also agree between GMB and $n_{Bulge}$ in there being no correlation. For the star-forming central galaxies, GMB has a decreasing $\sigma_{SFR}$ with halo mass while $n_{Bulge}$ shows no link between $\sigma_{SFR}$ and group mass. The opposite is seen for the star-forming satellies with GMB showing no relation and $n_{Bulge}$ a decreasing $\sigma_{SFR}$ with group mass. These differences may arise from GMB being less sensitive to the dust content of a galaxy than $n_{Bulge}$. Star-forming galaxies are known to be more dusty than quiescent galaxies \citep[e.g.][]{2014ApJ...782L..23H, 2019MNRAS.486.2827D, 2019ApJ...880..129P, 2020arXiv200809995D} thus, if GMB is less sensitive to dust, it is not surprising that GMB and $n_{Bulge}$ agree for the less dusty quiescent galaxies but disagree for the more dusy star-forming galaxies. It may also instead be influenced by galaxy interactions as, like dust, GMB is less sensitive to galaxy interactions than $n_{Bulge}$. Galaxy interactions are known to be able to trigger periods of star-formation \citep[e.g.][]{2013MNRAS.435.3627E, 2015MNRAS.454.1742K, 2018ApJ...868...46S, 2019A&A...631A..51P} and could drive the difference in trends seen between GMB and $n$.

The disagreement between $n_{Bulge}$ and GMB is less apparent in the range of these parameters, with GMB and $n_{Bulge}$ both showing no dependence of the range of parameter on the group mass for the star-forming galaxies and an increase in the range with group mass for the quiescent satellites. For the final galaxy sample, the quiescent central galaxies, GMB shows an increase in range with group mass while $n_{Bulge}$ shows no environmental dependence.

For the star-forming galaxies, the most bulge-like GMB galaxies (that is the highest GMB values) do not become more, or less, bulge-like as the group mass increases. This is also seen in $n_{Bulge}$ for the star-forming satellite galaxies, with the most bulge-like galaxies (that is the largest $n_{Bulge}$ values) not becoming more or less bulge-like as the mass increases. For the star-forming central galaxies, the maximum $n_{Bulge}$ increases from HB11 to HB12 before becoming constant. The most disk-like star-forming galaxies, those with the smallest GMB or $n_{Bulge}$, do not become more, or less, disk-like when examined with either $n_{Bulge}$ or GMB, showing that disk dominated galaxies are able to form stars in group environments, as would be expected. Thus, the star-forming galaxies' bulgieness is predominantly not impacted by the environment in which the galaxy lies.

For the quiescent galaxies, there is no agreement between GMB and $n_{Bulge}$ for the most and least bulge-like galaxies. GMB has the most bulge-like quiescent central galaxies not becoming more or less bulge-like while $n_{Bulge}$ sees these galaxies becoming more bulge-like. The least bulge-like of these same galaxies also have GMB showing no evolution while $n_{Bulge}$ becomes larger between HB11 and HB12 before becoming constant. For the quiescent satellite galaxies, GMB has the most bulge-like becoming more bulge-like as the group mass increases while the least bulge-like do not change. $n_{Bulge}$ on the other hand, shows the most bulge-like becomes less bulge-like in HB13 and the least bulge-like become less bulge-like as the group mass increases. As noted before, quiescent galaxies are less dusty than star-forming galaxies, so if dust is driving the difference between GMB and $n_{Bulge}$, then the quiescent galaxies should show better correlation between GMB and $n_{Bulge}$. As this is not the case, it is probable that dust is not what is causing the difference in results between GMB and $n_{Bulge}$.

Neither $n_{Bulge}$ or GMB show what is traditionally expected of galaxies in groups for the majority of the galaxies, that is they do not show an increase in bulge-likeness as the group mass increases. The exception is the most bulge-like quiescent satellites for GMB and the most bulge-like quiescent centrals for $n_{Bulge}$. The bulge S\'{e}rsic index also shows the opposite trend for the least bulge-like quiescent satellite galaxies, with both these galaxies becoming less bulge-like as the group mass increases.

The trends of the median parameter with sSFR do weakly agree between $n_{Bulge}$ and GMB. Both parameters show that all galaxies increase in disk-likeness an the sSFR increases, as would be expected. However, neither parameter shows as increase in bulge-likeness for satellite galaxies compared to central galaxies. The $n_{Bulge}$ also has no change in bulge-likeness between the central and satellite populations that is seen in GMB.

\subsection{Galaxy disruption}
To study galaxy disruption, we examine how asymmetry is affected by the environment. This allows the study of galaxies influenced by interactions and flybys as well as mergers \citep[e.g.][]{2016MNRAS.461.2589P}. Cuts of non-parametric statistics are not used as they are designed to select only merging galaxies \citep{2003ApJS..147....1C, 2004AJ....128..163L, 2008ApJ...672..177L} and not using these cuts also avoids the imperfect merger identification inherent in these classifications \citep{2019A&A...631A..51P}. As mentioned in Sect. \ref{subsec:disk-n-bulge}, mergers are expected to be less common as the group density increases \citep{2010ApJ...718.1158L, 2012A&A...539A..46A}. As a result, it would be expected that asymmetry should reflect this.

The maximum, minimum, and range of A should not be influenced by the fraction of galaxies that are interacting as these values are driven by the extreme cases, which should be independent of how often interactions occur. This is what is seen for the minimum for the majority of both central and satellite galaxies, the exceptions being the star-forming centrals and quiescent satellites in HB13. The maximum of the star-forming and quiescent central galaxies are also consistent with the field. However, the quiescent satellite galaxies show a decreasing maximum. This may possibly indicate a reduction in the number of quiescent satellite galaxies that are undergoing a merger, preventing higher values of asymmetry to be seen, or the more disruptive interactions are predominantly causing star-formation. The range of A is also lower in group environments than in the field for quiescent galaxies.

The standard deviations along M$_{\star}$ and SFR are all constant in group environments for the satellite galaxies and consistent with the field for all but the quiescent satellite galaxies, which have $\sigma_{SFR}$ that is larger. The central galaxies also have constant standard deviation along SFR that is consistent with the field. Along M$_{\star}$, the central galaxies are not constant but taking the variable changes with group mass as independence from group environment then the standard deviations of star-forming and quiescent central and satellite galaxies are not influenced by the group environment. However, the number of galaxy-galaxy interactions is expected to increase with group halo mass with fewer of these interactions resulting in a merger \citep{2012A&A...539A..46A}, which would result in more disrupted galaxies. This does not agree with our findings as such an increase in galaxy-galaxy interactions with halo mass would result in $\sigma_{M_{\star}}$ increasing with halo mass. There is, however, a weak indication of $\sigma_{SFR}$ increasing with halo mass for the quiescent galaxies, which would then agree with an increase in galaxy-galaxy interactions as the group mass increases.

The median of A appears to indicate that SFR for central galaxies, and to a lesser degree satellite galaxies, is driven by interactions. The increase in A at higher sSFR for group galaxies indicates that these higher SFRs are being caused by galaxy-galaxy interactions. Galaxy mergers and interactions are known to increase SFR of the interacting galaxies so this is not an unexpected result \citep[e.g.][]{2004MNRAS.350..798B, 2009ApJ...694L.123K, 2009PASJ...61..481S, 2019A&A...631A..51P}.

\subsection{Total S\'{e}rsic index and $r_{50}$}
For $n$ and $r_{50}$ it is possible to draw comparison to works looking at the distribution of $n$ and the effective radius ($r_{eff}$) from \citet{2011ApJ...742...96W} and \citet{2017MNRAS.465..619B}. While the $r_{50}$ studied here is not exactly the same as $r_{eff}$ - the latter being the radius within which 50\% of the light lies when a total S\'{e}rsic profile is fitted - the $r_{50}$ used in this work is not from a total S\'{e}rsic profile. While \citet{2011ApJ...742...96W} and \citet{2017MNRAS.465..619B} do not provide a fit for $n$ and $r_{eff}$, they do provide qualitative descriptions of the distributions.

\begin{figure}
	\resizebox{\hsize}{!}{\includegraphics{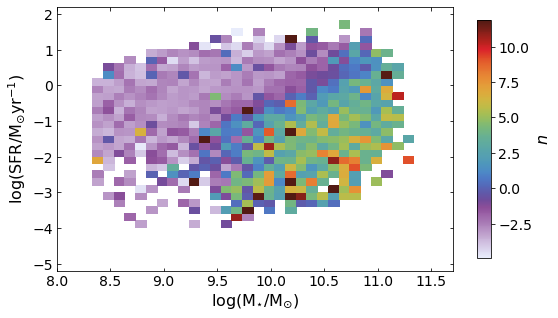}}
	\caption{Distribution of total S\'{e}rsic indices found in the field galaxies. As can be seen, $n$ depicts a MS in the SFR-M${\star}$ plane and, like \citet{2017MNRAS.465..619B}, we see no grouping of high $n$ above the MS.}
	\label{fig:sersic-plane}
\end{figure}

Both \citet{2011ApJ...742...96W} and \citet{2017MNRAS.465..619B} find that $n$ mimics the traditional MS derived from the number-density of galaxies in the SFR-M$_{\star}$ plane. This is also seen in \citet{2015ApJ...811L..12W}. The differences arise between these studies with \citet{2017MNRAS.465..619B} not seeing a population of high $n$ as galaxies move above the MS. In this work, we also see a MS depicted by $n$ in all environments as the example for field galaxies in Fig. \ref{fig:sersic-plane} shows. As with \citet{2017MNRAS.465..619B}, we do not see a population of high $n$ galaxies lying above the MS. \citet{2017MNRAS.465..619B} claims that the lack of the high $n$ population is a result of their selection criteria, which removes galaxies that are not well fit by a total S\'{e}rsic profile and is likely to be removing star-bursting galaxies with high $n$ that are likely to be merging. However, in this work we do not make such a cut and yet find similar results. At low redshifts comparable to this work, \citet{2017MNRAS.465..619B} use galaxies in GAMA, as we do here, but derive their own $n$ so this lack of difference is not unexpected.

For $r_{eff}$, \citet{2011ApJ...742...96W} and \citet{2017MNRAS.465..619B} find that galaxies on the MS have larger radii than galaxies off the MS (both above and below) at the same M$_{\star}$. We also see such a trend with $r_{50}$: Galaxies on the MS have a larger radii than those above and below at the same M$_{\star}$.

\section{Summary and conclusions}\label{sec:conc}
The environment in which a galaxy lies is know to influence the morphology of that galaxy. In this work, we aimed to examine exactly how the environment influences the concentration (C), Gini, the second-order moment of the brightest 20\% of the light (M$_{20}$), asymmetry (A), Gini-M$_{20}$ bulge (GMB) non-parametric statistics as well as the 50\% light radius ($r_{50}$), total S\'{e}rsic index ($n$) and bulge S\'{e}rsic index ($n_{Bulge}$) parametric statistics for galaxies. This was done for a sample of GAMA galaxies with known environment (field or group) and with these parameters derived using optical r-band KiDS imaging. So we could study the differences between low and high mass group environments, the group galaxy sample was further divided into bins with halo masses (M$_{h}$) between $11 \leq \mathrm{log(M}_{h}/\mathrm{M}_\odot) < 12$ (HB11), $12 \leq \mathrm{log(M}_{h}/\mathrm{M}_\odot) < 13$ (HB12), and $13 \leq \mathrm{log(M}_{h}/\mathrm{M}_\odot) < 14$ (HB13) with the central and satellite galaxies within each bin separated from one another. We then explored how the different parameters are distributed across the SFR-M$_{\star}$ plane in the differing environments.

Using our simple split between star-forming and quiescent galaxies, we find that the fraction of star-forming galaxies decreases as the environment becomes more massive, which is qualitatively consistent with other works. We see that 79.1\% of field galaxies are star-forming, dropping to 72.6\% in HB11, 62.0\% in HB12, and 48.7\% in the most massive environments. When we also split for central and satellite galaxies in the groups, we find that 69.2\%, 51.2\%, and 40.1\% of central galaxies are star-forming in HB11, HB12, and HB13, respectively, while 75.4\%, 68.0\%, and 51.1\% of satellite galaxies are forming stars. Thus a greater fraction of satellite galaxies are star-forming compared to centrals.

As expected, we also find that galaxy environments cause some change to the distributions of parametric and non-parametric morphological indicators. The compactness of a galaxy, determined with C, Gini, M$_{20}$, $r_{50}$ and $n$, does show variation with group halo mass for the star-forming and quiescent central and satellite galaxies. However, how this changes depends on the indicator being examined, with all five parameters showing slightly differing trends. There is an indication that star-forming satellite galaxies and all quiescent galaxies have compactness that is independent of M$_{\star}$. The impact on the compactness as SFR changes is less clear, with different parameters showing different trends for the different galaxy sample. Using C and $r_{50}$, we see that all galaxies become larger as the group halo mass increases, for both central and satellite galaxies. We also see that the size of the largest galaxies in the low mass group halos increases faster than the size of the smallest galaxies as the environment mass increases.

For the bulge or disk domination, there is little agreement between GMB and $n_{Bulge}$. Where there is agreement we see that the maximum, minimum, and range of these two parameters for the star-forming galaxies do not change as the group halo mass increases, implying that the most bulge-like and most disk-like galaxies do not become more or less bulge-like or disk-like as the environment changes. The quiescent galaxies most and least bulge-like galaxies have differing tends when studying GMB or $n_{Bulge}$. Where GMB and $n_{Bulge}$ do agree is that the range of parameter for the quiescent satellites does increase with group halo mass but GMB finds the most bulge-like galaxies driving this change while $n_{Bulge}$ finds the change caused by the most disk-like. GMB and $n_{Bulge}$ also only agree that the variability of disk or bulge domination as M$_{\star}$ changes is independent of group environment for the satellite galaxies and that the variability of bulge or disk domination along SFR is also independent of group halo mass for the quiescent satellites.

We find little evidence for more galaxy disruption as the environment mass increases. Using A, we see no evidence for an increase in the disruption of galaxies at a fixed SFR or M$_{\star}$ as the group halo mass increases. We also see that the least disrupted galaxies, that is the galaxies with the lowest A, and the most disrupted galaxies are predominantly uninfluenced by the environment. The exception being the quiescent satellite galaxies that see a reduction in A as the halo mass increases. This lack of environmental dependence was unexpected as there is known to be an increase in galaxy-galaxy interactions in higher mass groups, despite a decrease in interactions that result in a merger \citep{2010ApJ...718.1158L, 2012A&A...539A..46A}. With higher interaction rates, it would be expected to see more disruption of the galaxy population at fixed SFR or M$_{\star}$. The most and least disturbed galaxies would be expected to remain constant as these are driven by the extreme cases that should be interaction-rate independent in large galaxy samples.

The trends found in this work are weak, a result of the relatively small sample sizes for the group galaxies, with only 900 central galaxies in HB13. Larger sample sizes would enable the fitting coefficients to be better constrained and make any trends in the morphological parameters clearer. The work presented here also looks at galaxies that are at a relatively low redshift (z < 0.15). Thus we do not currently examine how, or indeed if, these changes to morphology in the SFR-M$_{\star}$ plane due to environment change as we look further back. It would therefore be informative to perform a similar study at higher redshifts. This would require high resolution imaging over a large area of the sky to be able to derive the morphological parameters from the image plane for a statistically large sample of galaxies in a range of environments, for example with the Vera C. Rubin Observatory \citep{2019NatRP...1..450R}.

%One column figure
%\begin{figure}
%	\resizebox{\hsize}{!}{\includegraphics{}}
%	\caption{<Your caption text...>.}
%	\label{<Your label>}
%\end{figure}

\begin{acknowledgements}
We would like to thank the anonymous referee for their thoughtful comments that have improved the quality of this paper.
	
We would like to thank A.~Graham for helpful discussions on this paper.

W.J.P. has been supported by the Polish National Science Center project UMO-2018/30/M/ST9/00757.

GAMA is a joint European-Australasian project based around a spectroscopic campaign using the Anglo-Australian Telescope. The GAMA input catalogue is based on data taken from the Sloan Digital Sky Survey and the UKIRT Infrared Deep Sky Survey. Complementary imaging of the GAMA regions is being obtained by a number of independent survey programmes including GALEX MIS, VST KiDS, VISTA VIKING, WISE, Herschel-ATLAS, GMRT and ASKAP providing UV to radio coverage. GAMA is funded by the STFC (UK), the ARC (Australia), the AAO, and the participating institutions. The GAMA website is \url{http://www.gama-survey.org/} .

Based on observations made with ESO Telescopes at the La Silla Paranal Observatory under programme IDs 177.A-3016, 177.A-3017, 177.A-3018 and 179.A-2004, and on data products produced by the KiDS consortium. The KiDS production team acknowledges support from: Deutsche Forschungsgemeinschaft, ERC, NOVA and NWO-M grants; Target; the University of Padova, and the University Federico II (Naples).
\end{acknowledgements}

\bibliographystyle{aa} % style aa.bst
\bibliography{GAMA-405-morphMS} % References as <filename>.bib

\begin{appendix}

\section{Example corner plot}\label{app:corner}
\begin{figure*}
	\centering
	\resizebox{1.0\hsize}{!}{\includegraphics{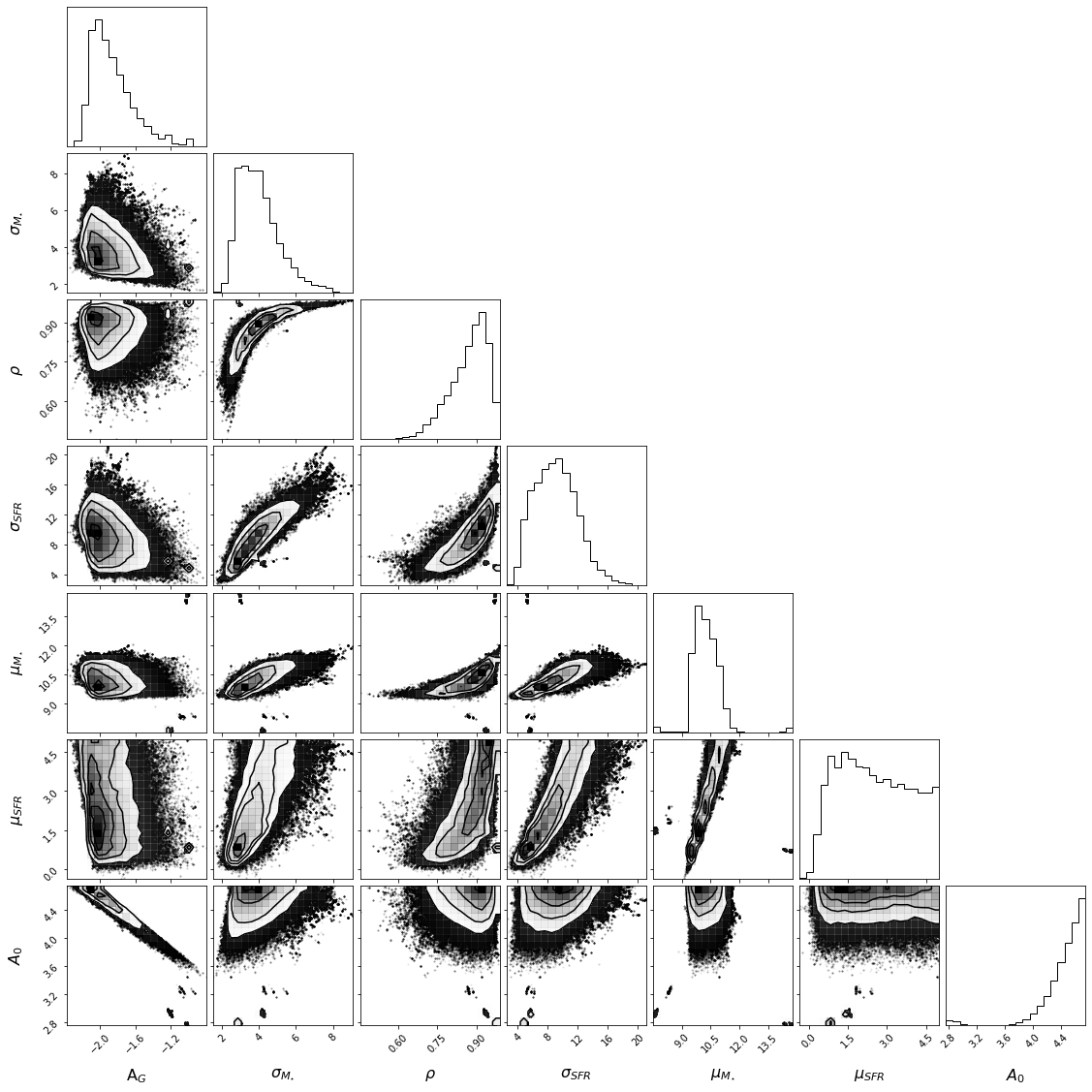}}
	\caption{Example corner plot for fitting the concentration of the star-forming field galaxies.}
	\label{fig:app:corner}
\end{figure*}

Here we present an example corner plot \citep{corner} for the star-forming field galaxies' coefficients when fitting concentration in Fig. \ref{fig:app:corner}. In this example, the normalisation ($A_{0}$) and amplitude of the Gaussian ($A_{G}$) are highly correlated, as can be seen. While the posterior distribution of $A_{0}$ is high at the end of the prior range, if this range is extended such that the posterior peaks within the prior range, the posterior of $A_{G}$ then peaks at the edge of its prior range. Expanding both prior ranges results in $A_{G}$ and $A_{0}$ that are unrealistic while not notably increasing the goodness of fit. As such, it was deemed acceptable to not make the priors less restrictive. This issue was not seen for other morphological parameters. The two means, $\mu_{M_{\star}}$ and $\mu_{SFR}$, are also highly correlated in this example and the two standard deviations also show some correlation as well as $\rho$ and $\sigma_{SFR}$. The other parameter pairs appear to be predominantly uncorrelated.

\section{Fitting coefficients}\label{app:coefficients}
Here we present the best fit coefficients for all morphological parameters in all environments. The coefficients for the star-forming galaxies, as defined by Eq. \ref{eq:sfcut}, are in Table \ref{tbl:app:sf} while the coefficients for the quiescent galaxies are in Table \ref{tbl:app:qu}.

\setlength{\tabcolsep}{3pt}

%\begin{landscape}
\begin{table*}
	\caption{Coefficients from fitting a single Gaussian distribution (Eq. \ref{eq:single}) to morphological parameters for the star-forming galaxies as defined by Eq. \ref{eq:sfcut}. Env. is the environment, Cen. and Sat. are the central and satellite galaxies, respectively, and Prm. is the parameter being studied.}
	\label{tbl:app:sf}
	\centering
	\begin{tabular}{ccccccccc}
		\hline
		Env. & Prm. & $A_{0}$ & $A_{G}$ & $\sigma_{M_{\star}}$ & $\rho$ & $\sigma_{SFR}$ & $\mu_{M_{\star}}$ & $\mu_{SFR}$ \\
		\hline
		\multirow{8}{*}{Field}
		& C & 4.721 $\pm$ 0.218 & -2.081 $\pm$ 0.219 & 3.395 $\pm$ 1.051 & 0.833 $\pm$ 0.072 & 6.646 $\pm$ 2.921 & 9.787 $\pm$ 0.581 & 1.159 $\pm$ 1.558 \\
		& Gini & 0.862 $\pm$ 0.064 & -0.371 $\pm$ 0.059 & 2.968 $\pm$ 0.379 & 0.621 $\pm$ 0.080 & 4.888 $\pm$ 1.345 & 9.583 $\pm$ 0.165 & 0.537 $\pm$ 0.528 \\
		& M$_{20}$ & -3.449 $\pm$ 0.483 & 1.847 $\pm$ 0.480 & 7.486 $\pm$ 3.186 & 0.909 $\pm$ 0.167 & 20.673 $\pm$ 7.541 & 10.139 $\pm$ 0.863 & 3.423 $\pm$ 2.589 \\
		& A & -0.028 $\pm$ 0.023 & 0.144 $\pm$ 0.028 & 3.862 $\pm$ 8.485 & 0.554 $\pm$ 0.730 & 0.912 $\pm$ 0.811 & 12.255 $\pm$ 2.618 & 1.313 $\pm$ 0.578 \\
		& GMB & 0.339 $\pm$ 0.050 & -0.740 $\pm$ 0.038 & 1.582 $\pm$ 0.420 & 0.707 $\pm$ 0.109 & 2.877 $\pm$ 1.405 & 9.512 $\pm$ 0.340 & 0.551 $\pm$ 0.999 \\
		& $r_{50}$ & 2.169 $\pm$ 0.004 & 17.998 $\pm$ 0.278 & 11.409 $\pm$ 1.616 & -0.961 $\pm$ 0.021 & 12.879 $\pm$ 1.786 & 11.999 $\pm$ 0.004 & 4.999 $\pm$ 0.005 \\
		& $n$ & 3.306 $\pm$ 0.016 & -2.180 $\pm$ 0.021 & 1.980 $\pm$ 0.082 & 0.828 $\pm$ 0.015 & 1.945 $\pm$ 0.083 & 9.030 $\pm$ 0.045 & -0.303 $\pm$ 0.042 \\
		& $n_{Bulge}$ & 6.381 $\pm$ 0.017 & -5.304 $\pm$ 0.063 & 1.464 $\pm$ 0.527 & -0.901 $\pm$ 0.035 & 1.872 $\pm$ 0.682 & 8.435 $\pm$ 0.192 & -0.767 $\pm$ 0.237 \\
		\hline
		\multirow{8}{*}{\shortstack{HB11 \\ Cen.}}
		& C & 4.346 $\pm$ 0.254 & -1.691 $\pm$ 0.271 & 2.813 $\pm$ 4.583 & 0.885 $\pm$ 0.065 & 4.569 $\pm$ 8.675 & 9.536 $\pm$ 1.319 & 0.377 $\pm$ 2.812 \\
		& Gini & 0.756 $\pm$ 0.209 & -0.258 $\pm$ 0.364 & 7.854 $\pm$ 10.376 & 0.920 $\pm$ 0.672 & 19.857 $\pm$ 10.521 & 9.333 $\pm$ 3.999 & 1.979 $\pm$ 5.274 \\
		& M$_{20}$ & -3.275 $\pm$ 0.556 & 1.710 $\pm$ 0.538 & 2.924 $\pm$ 3.105 & 0.722 $\pm$ 0.575 & 6.021 $\pm$ 8.590 & 9.613 $\pm$ 1.647 & 1.047 $\pm$ 4.361 \\
		& A & -0.038 $\pm$ 0.030 & 0.202 $\pm$ 0.046 & 18.325 $\pm$ 9.540 & 0.832 $\pm$ 0.615 & 2.980 $\pm$ 8.870 & 12.332 $\pm$ 2.692 & 2.262 $\pm$ 4.910 \\
		& GMB & 0.215 $\pm$ 0.086 & -0.556 $\pm$ 0.074 & 5.965 $\pm$ 1.540 & 0.992 $\pm$ 0.619 & 11.724 $\pm$ 8.877 & 9.287 $\pm$ 0.618 & -0.165 $\pm$ 4.035 \\
		& $r_{50}$ & 2.342 $\pm$ 0.257 & 5.677 $\pm$ 0.557 & 1.889 $\pm$ 0.289 & -0.106 $\pm$ 0.231 & 1.227 $\pm$ 0.381 & 11.970 $\pm$ 0.180 & 1.202 $\pm$ 0.299 \\
		& $n$ & 3.616 $\pm$ 0.117 & -2.769 $\pm$ 0.150 & 4.702 $\pm$ 2.736 & 0.980 $\pm$ 0.017 & 7.547 $\pm$ 5.509 & 7.424 $\pm$ 0.759 & -3.157 $\pm$ 1.324 \\
		& $n_{Bulge}$ & 4.980 $\pm$ 0.065 & -6.238 $\pm$ 1.369 & 0.519 $\pm$ 0.152 & -0.584 $\pm$ 0.460 & 8.683 $\pm$ 6.995 & 9.348 $\pm$ 0.179 & -9.088 $\pm$ 3.007 \\
		\hline
		\multirow{8}{*}{\shortstack{HB11 \\ Sat.}}
		& C & 2.631 $\pm$ 0.949 & 0.846 $\pm$ 0.892 & 10.967 $\pm$ 5.711 & 0.992 $\pm$ 0.380 & 18.922 $\pm$ 7.980 & 14.223 $\pm$ 1.693 & 3.568 $\pm$ 4.339 \\
		& Gini & 0.811 $\pm$ 0.208 & -0.311 $\pm$ 0.353 & 15.458 $\pm$ 10.024 & 0.968 $\pm$ 0.679 & 16.753 $\pm$ 10.087 & 9.686 $\pm$ 4.114 & 1.162 $\pm$ 5.127 \\
		& M$_{20}$ & -1.972 $\pm$ 0.711 & 0.395 $\pm$ 0.684 & 2.435 $\pm$ 9.376 & 0.429 $\pm$ 0.632 & 3.045 $\pm$ 8.915 & 8.191 $\pm$ 2.971 & 0.383 $\pm$ 4.905 \\
		& A & -0.004 $\pm$ 0.053 & 0.126 $\pm$ 0.058 & 11.623 $\pm$ 9.335 & -0.923 $\pm$ 0.628 & 3.854 $\pm$ 8.923 & 13.464 $\pm$ 2.708 & 0.783 $\pm$ 4.850 \\
		& GMB & -0.323 $\pm$ 0.194 & 0.302 $\pm$ 0.301 & 1.389 $\pm$ 10.624 & 0.972 $\pm$ 0.731 & 1.711 $\pm$ 9.842 & 11.155 $\pm$ 2.512 & 0.705 $\pm$ 4.919 \\
		& $r_{50}$ & 1.524 $\pm$ 0.157 & 5.065 $\pm$ 0.268 & 2.147 $\pm$ 0.117 & 0.535 $\pm$ 0.084 & 1.373 $\pm$ 0.111 & 11.985 $\pm$ 0.108 & 1.356 $\pm$ 0.160 \\
		& $n$ & 3.359 $\pm$ 0.302 & -2.277 $\pm$ 0.299 & 2.254 $\pm$ 1.566 & 0.842 $\pm$ 0.063 & 1.754 $\pm$ 1.140 & 9.068 $\pm$ 0.519 & -0.024 $\pm$ 0.357 \\
		& $n_{Bulge}$ & 6.336 $\pm$ 0.105 & -7.815 $\pm$ 0.998 & 0.985 $\pm$ 0.127 & -0.510 $\pm$ 0.082 & 1.411 $\pm$ 0.167 & 7.582 $\pm$ 0.247 & -0.072 $\pm$ 0.182 \\
		\hline
		\multirow{8}{*}{\shortstack{HB12 \\ Cen.}}
		& C & 3.874 $\pm$ 0.352 & -1.228 $\pm$ 0.348 & 3.490 $\pm$ 3.901 & 0.960 $\pm$ 0.044 & 5.841 $\pm$ 8.098 & 8.832 $\pm$ 1.380 & -1.123 $\pm$ 2.911 \\
		& Gini & 0.769 $\pm$ 0.203 & -0.363 $\pm$ 0.345 & 13.397 $\pm$ 10.163 & 0.896 $\pm$ 0.664 & 13.484 $\pm$ 10.205 & 1.377 $\pm$ 4.151 & -4.275 $\pm$ 5.229 \\
		& M$_{20}$ & -2.723 $\pm$ 0.608 & 1.167 $\pm$ 0.567 & 8.227 $\pm$ 3.121 & 0.972 $\pm$ 0.485 & 13.352 $\pm$ 8.226 & 6.316 $\pm$ 1.573 & -4.735 $\pm$ 3.943 \\
		& A & -0.020 $\pm$ 0.055 & 0.185 $\pm$ 0.057 & 13.040 $\pm$ 9.205 & 0.911 $\pm$ 0.621 & 3.891 $\pm$ 8.895 & 11.986 $\pm$ 2.694 & 2.789 $\pm$ 4.840 \\
		& GMB & -0.277 $\pm$ 0.185 & 0.384 $\pm$ 0.378 & 3.310 $\pm$ 3.136 & 0.994 $\pm$ 0.496 & 5.173 $\pm$ 8.914 & 10.795 $\pm$ 1.135 & 0.099 $\pm$ 3.666 \\
		& $r_{50}$ & 3.622 $\pm$ 0.052 & 6.952 $\pm$ 0.330 & 1.158 $\pm$ 0.049 & -0.213 $\pm$ 0.117 & 0.866 $\pm$ 0.082 & 11.987 $\pm$ 0.050 & 0.796 $\pm$ 0.100 \\
		& $n$ & 3.639 $\pm$ 0.277 & -2.785 $\pm$ 0.311 & 3.717 $\pm$ 1.163 & 0.967 $\pm$ 0.021 & 5.069 $\pm$ 2.135 & 7.725 $\pm$ 0.729 & -2.329 $\pm$ 1.047 \\
		& $n_{Bulge}$ & 5.448 $\pm$ 0.053 & -12.085 $\pm$ 2.576 & 0.689 $\pm$ 0.044 & 0.986 $\pm$ 0.051 & 1.363 $\pm$ 0.117 & 8.365 $\pm$ 0.089 & -2.102 $\pm$ 0.261 \\
		\hline
		\multirow{8}{*}{\shortstack{HB12 \\ Sat.}}
		& C & 4.213 $\pm$ 0.323 & -1.581 $\pm$ 0.318 & 2.000 $\pm$ 1.120 & 0.357 $\pm$ 0.460 & 4.303 $\pm$ 6.140 & 9.328 $\pm$ 0.578 & 1.022 $\pm$ 1.800 \\
		& Gini & 0.754 $\pm$ 0.201 & -0.256 $\pm$ 0.341 & 6.413 $\pm$ 10.076 & 0.892 $\pm$ 0.681 & 19.060 $\pm$ 10.133 & 9.592 $\pm$ 4.216 & 1.687 $\pm$ 5.295 \\
		& M$_{20}$ & -2.976 $\pm$ 0.622 & 1.421 $\pm$ 0.617 & 4.790 $\pm$ 5.472 & 0.808 $\pm$ 0.583 & 11.467 $\pm$ 7.950 & 10.151 $\pm$ 1.697 & 3.530 $\pm$ 3.789 \\
		& A & -0.034 $\pm$ 0.049 & 0.129 $\pm$ 0.055 & 7.268 $\pm$ 9.748 & -0.509 $\pm$ 0.660 & 1.048 $\pm$ 10.014 & 14.412 $\pm$ 2.596 & 0.627 $\pm$ 4.887 \\
		& GMB & -0.329 $\pm$ 0.178 & 0.367 $\pm$ 0.366 & 1.313 $\pm$ 1.406 & 0.962 $\pm$ 0.603 & 2.359 $\pm$ 9.763 & 10.930 $\pm$ 1.149 & 0.464 $\pm$ 4.209 \\
		& $r_{50}$ & 2.269 $\pm$ 0.071 & 4.469 $\pm$ 0.146 & 2.247 $\pm$ 0.132 & -0.306 $\pm$ 0.100 & 1.255 $\pm$ 0.124 & 11.997 $\pm$ 0.036 & 0.631 $\pm$ 0.092 \\
		& $n$ & 3.204 $\pm$ 0.017 & -1.912 $\pm$ 0.037 & 11.268 $\pm$ 5.510 & 0.996 $\pm$ 0.004 & 12.539 $\pm$ 6.345 & 7.639 $\pm$ 0.759 & -1.873 $\pm$ 0.812 \\
		& $n_{Bulge}$ & 6.713 $\pm$ 0.047 & -6.847 $\pm$ 0.439 & 1.659 $\pm$ 0.094 & -0.845 $\pm$ 0.023 & 1.286 $\pm$ 0.099 & 7.397 $\pm$ 0.141 & -0.001 $\pm$ 0.068 \\
		\hline
		\multirow{8}{*}{\shortstack{HB13 \\ Cen.}}
		& C & 4.476 $\pm$ 0.338 & -1.903 $\pm$ 0.542 & 3.290 $\pm$ 3.372 & 0.914 $\pm$ 0.297 & 4.339 $\pm$ 9.049 & 8.752 $\pm$ 1.167 & -0.960 $\pm$ 3.372 \\
		& Gini & 0.744 $\pm$ 0.202 & -0.281 $\pm$ 0.361 & 11.327 $\pm$ 10.030 & 0.916 $\pm$ 0.684 & 15.271 $\pm$ 10.308 & 5.451 $\pm$ 4.162 & -2.281 $\pm$ 5.149 \\
		& M$_{20}$ & -3.231 $\pm$ 0.643 & 1.710 $\pm$ 0.637 & 4.237 $\pm$ 5.304 & 0.790 $\pm$ 0.695 & 8.447 $\pm$ 8.295 & 9.729 $\pm$ 1.818 & 2.148 $\pm$ 4.968 \\
		& A & -0.071 $\pm$ 0.049 & 0.212 $\pm$ 0.056 & 24.802 $\pm$ 9.288 & -0.866 $\pm$ 0.629 & 5.842 $\pm$ 8.965 & 14.677 $\pm$ 2.660 & 2.122 $\pm$ 4.902 \\
		& GMB & 0.254 $\pm$ 0.281 & -0.641 $\pm$ 0.550 & 2.288 $\pm$ 1.247 & 0.949 $\pm$ 0.779 & 5.187 $\pm$ 8.859 & 8.944 $\pm$ 1.222 & -1.397 $\pm$ 4.569 \\
		& $r_{50}$ & 3.486 $\pm$ 0.062 & 10.073 $\pm$ 0.454 & 1.077 $\pm$ 0.040 & 0.075 $\pm$ 0.089 & 0.965 $\pm$ 0.077 & 11.998 $\pm$ 0.019 & 1.257 $\pm$ 0.107 \\
		& $n$ & 4.836 $\pm$ 0.438 & -3.854 $\pm$ 0.464 & 1.570 $\pm$ 1.386 & 0.737 $\pm$ 0.091 & 1.974 $\pm$ 2.862 & 9.502 $\pm$ 0.688 & -0.019 $\pm$ 1.100 \\
		& $n_{Bulge}$ & 5.422 $\pm$ 0.134 & -7.591 $\pm$ 2.767 & 3.483 $\pm$ 1.411 & -0.993 $\pm$ 0.948 & 3.132 $\pm$ 1.363 & 14.937 $\pm$ 1.164 & -2.898 $\pm$ 0.991 \\
		\hline
		\multirow{8}{*}{\shortstack{HB13 \\ Sat.}}
		& C & 4.944 $\pm$ 0.494 & -2.321 $\pm$ 0.492 & 2.071 $\pm$ 1.090 & 0.397 $\pm$ 0.519 & 5.979 $\pm$ 8.335 & 9.543 $\pm$ 0.453 & 1.239 $\pm$ 2.936 \\
		& Gini & 0.794 $\pm$ 0.214 & -0.368 $\pm$ 0.324 & 4.091 $\pm$ 10.018 & -0.217 $\pm$ 0.659 & 8.063 $\pm$ 10.153 & 8.045 $\pm$ 4.143 & 4.813 $\pm$ 5.180 \\
		& M$_{20}$ & -3.664 $\pm$ 0.625 & 2.117 $\pm$ 0.599 & 3.055 $\pm$ 3.032 & 0.392 $\pm$ 0.587 & 13.000 $\pm$ 8.188 & 9.577 $\pm$ 1.652 & 3.676 $\pm$ 4.429 \\
		& A & -0.015 $\pm$ 0.040 & 0.122 $\pm$ 0.045 & 8.877 $\pm$ 9.511 & 0.340 $\pm$ 0.642 & 1.000 $\pm$ 9.365 & 11.260 $\pm$ 2.604 & 1.311 $\pm$ 4.907 \\
		& GMB & -0.321 $\pm$ 0.173 & 0.389 $\pm$ 0.370 & 1.661 $\pm$ 0.986 & 0.972 $\pm$ 0.678 & 4.813 $\pm$ 8.995 & 10.719 $\pm$ 0.987 & -0.345 $\pm$ 4.049 \\
		& $r_{50}$ & 2.943 $\pm$ 0.053 & 7.613 $\pm$ 0.263 & 1.193 $\pm$ 0.031 & 0.451 $\pm$ 0.054 & 0.670 $\pm$ 0.030 & 11.999 $\pm$ 0.024 & 0.838 $\pm$ 0.053 \\
		& $n$ & 2.760 $\pm$ 0.062 & -1.675 $\pm$ 0.072 & 10.064 $\pm$ 3.325 & 0.999 $\pm$ 0.006 & 24.088 $\pm$ 8.122 & 9.010 $\pm$ 0.759 & -0.728 $\pm$ 1.786 \\
		& $n_{Bulge}$ & 5.411 $\pm$ 0.038 & -11.463 $\pm$ 2.214 & 1.286 $\pm$ 0.167 & -0.839 $\pm$ 0.075 & 0.700 $\pm$ 0.101 & 7.049 $\pm$ 0.389 & -0.099 $\pm$ 0.199 \\
		\hline
	\end{tabular}
\end{table*}
%\end{landscape}

%\begin{landscape}
\begin{table*}
	\caption{Coefficients from fitting a single Gaussian distribution (Eq. \ref{eq:single}) to morphological parameters for the quiescent galaxies as defined by Eq. \ref{eq:sfcut}. Env. is the environment, Cen. and Sat. are the central and satellite galaxies, respectively, and Prm. is the parameter being studied.}
	\label{tbl:app:qu}
	\centering
	\begin{tabular}{ccccccccc}
		\hline
		Env. & Prm. & $A_{0}$ & $A_{G}$ & $\sigma_{M_{\star}}$ & $\rho$ & $\sigma_{SFR}$ & $\mu_{M_{\star}}$ & $\mu_{SFR}$ \\
		\hline
		\multirow{8}{*}{Field}
		& C & 2.841 $\pm$ 0.092 & 0.799 $\pm$ 1.036 & 0.763 $\pm$ 0.612 & 0.361 $\pm$ 0.393 & 9.187 $\pm$ 2.118 & 11.047 $\pm$ 1.394 & -5.744 $\pm$ 4.374 \\
		& Gini & 0.688 $\pm$ 0.011 & -0.179 $\pm$ 0.011 & 1.967 $\pm$ 0.090 & 0.047 $\pm$ 0.039 & 9.998 $\pm$ 0.177 & 9.052 $\pm$ 0.027 & 0.533 $\pm$ 0.332 \\
		& M$_{20}$ & -2.439 $\pm$ 0.502 & 0.771 $\pm$ 0.521 & 2.077 $\pm$ 1.381 & 0.335 $\pm$ 0.382 & 9.752 $\pm$ 1.445 & 8.689 $\pm$ 1.303 & -1.876 $\pm$ 2.903 \\
		& A & -0.018 $\pm$ 0.000 & 0.110 $\pm$ 0.000 & 1.571 $\pm$ 0.002 & 0.746 $\pm$ 0.001 & 3.843 $\pm$ 0.005 & 12.589 $\pm$ 0.002 & 3.895 $\pm$ 0.006 \\
		& GMB & -0.261 $\pm$ 0.053 & 0.965 $\pm$ 0.137 & 1.587 $\pm$ 0.282 & 0.769 $\pm$ 0.324 & 8.094 $\pm$ 2.001 & 12.526 $\pm$ 0.422 & 4.362 $\pm$ 3.334 \\
		& $r_{50}$ & 2.107 $\pm$ 0.002 & 13.575 $\pm$ 0.530 & 1.403 $\pm$ 0.069 & -0.652 $\pm$ 0.033 & 9.954 $\pm$ 0.288 & 12.000 $\pm$ 0.010 & 4.975 $\pm$ 0.130 \\
		& $n$ & 3.248 $\pm$ 0.021 & -2.179 $\pm$ 0.182 & 1.258 $\pm$ 0.197 & 0.889 $\pm$ 0.063 & 9.953 $\pm$ 1.376 & 8.413 $\pm$ 0.277 & -6.051 $\pm$ 2.206 \\
		& $n_{Bulge}$ & 6.758 $\pm$ 0.026 & -5.776 $\pm$ 0.159 & 0.429 $\pm$ 0.020 & 0.192 $\pm$ 0.165 & 5.318 $\pm$ 0.801 & 8.752 $\pm$ 0.026 & -0.022 $\pm$ 0.307 \\
		\hline
		\multirow{8}{*}{\shortstack{HB11 \\ Cen.}}
		& C & 2.945 $\pm$ 1.284 & 0.821 $\pm$ 2.154 & 1.912 $\pm$ 2.092 & 0.861 $\pm$ 0.573 & 12.296 $\pm$ 7.932 & 11.491 $\pm$ 2.628 & -1.084 $\pm$ 4.822 \\
		& Gini & 0.664 $\pm$ 0.238 & -0.405 $\pm$ 0.390 & 6.917 $\pm$ 10.061 & 0.772 $\pm$ 0.670 & 19.428 $\pm$ 10.268 & 4.765 $\pm$ 4.045 & 4.444 $\pm$ 5.195 \\
		& M$_{20}$ & -2.139 $\pm$ 0.756 & 1.288 $\pm$ 0.777 & 5.082 $\pm$ 10.585 & 0.921 $\pm$ 0.628 & 25.968 $\pm$ 9.812 & 5.825 $\pm$ 3.515 & -4.671 $\pm$ 5.182 \\
		& A & -0.001 $\pm$ 0.029 & 0.059 $\pm$ 0.030 & 7.892 $\pm$ 9.910 & -0.765 $\pm$ 0.663 & 4.046 $\pm$ 9.776 & 9.315 $\pm$ 2.695 & 2.970 $\pm$ 4.949 \\
		& GMB & -0.220 $\pm$ 0.335 & 0.605 $\pm$ 0.509 & 0.764 $\pm$ 10.626 & 0.382 $\pm$ 0.696 & 3.117 $\pm$ 9.182 & 10.919 $\pm$ 2.638 & -2.295 $\pm$ 5.102 \\
		& $r_{50}$ & 1.761 $\pm$ 0.322 & 6.128 $\pm$ 1.583 & 2.592 $\pm$ 1.212 & -0.913 $\pm$ 0.145 & 11.596 $\pm$ 5.995 & 11.932 $\pm$ 0.349 & -0.698 $\pm$ 2.123 \\
		& $n$ & 3.792 $\pm$ 0.100 & -4.450 $\pm$ 0.946 & 2.043 $\pm$ 2.856 & 0.946 $\pm$ 0.057 & 4.077 $\pm$ 8.449 & 7.167 $\pm$ 0.760 & -5.289 $\pm$ 1.914 \\
		& $n_{Bulge}$ & 6.544 $\pm$ 0.083 & -8.302 $\pm$ 2.638 & 0.458 $\pm$ 0.238 & -0.919 $\pm$ 0.565 & 0.720 $\pm$ 12.408 & 11.334 $\pm$ 0.226 & -2.476 $\pm$ 3.001 \\
		\hline
		\multirow{8}{*}{\shortstack{HB11 \\ Sat.}}
		& C & 2.571 $\pm$ 1.115 & 1.616 $\pm$ 1.094 & 3.460 $\pm$ 4.023 & 0.936 $\pm$ 0.615 & 3.891 $\pm$ 8.693 & 14.805 $\pm$ 1.814 & 2.964 $\pm$ 4.676 \\
		& Gini & 0.835 $\pm$ 0.225 & -0.316 $\pm$ 0.368 & 4.974 $\pm$ 10.291 & 0.039 $\pm$ 0.683 & 8.692 $\pm$ 10.445 & 8.248 $\pm$ 4.020 & -0.510 $\pm$ 5.311 \\
		& M$_{20}$ & -2.906 $\pm$ 0.716 & 1.401 $\pm$ 0.696 & 5.956 $\pm$ 9.822 & 0.395 $\pm$ 0.494 & 24.350 $\pm$ 10.257 & 6.342 $\pm$ 3.084 & 0.629 $\pm$ 5.294 \\
		& A & -0.047 $\pm$ 0.039 & 0.081 $\pm$ 0.036 & 4.905 $\pm$ 9.997 & 0.978 $\pm$ 0.656 & 16.238 $\pm$ 9.763 & 8.431 $\pm$ 2.668 & -8.948 $\pm$ 4.965 \\
		& GMB & 0.198 $\pm$ 0.181 & -0.351 $\pm$ 0.258 & 1.281 $\pm$ 10.601 & 0.384 $\pm$ 0.699 & 3.250 $\pm$ 10.043 & 8.701 $\pm$ 2.708 & -1.389 $\pm$ 5.017 \\
		& $r_{50}$ & 0.865 $\pm$ 0.457 & 6.609 $\pm$ 0.784 & 3.367 $\pm$ 1.635 & -0.537 $\pm$ 0.274 & 8.808 $\pm$ 4.192 & 11.961 $\pm$ 0.477 & 4.997 $\pm$ 1.051 \\
		& $n$ & 5.484 $\pm$ 0.182 & -4.138 $\pm$ 0.301 & 3.525 $\pm$ 0.517 & -0.902 $\pm$ 0.041 & 2.357 $\pm$ 0.408 & 7.094 $\pm$ 0.533 & -0.436 $\pm$ 0.339 \\
		& $n_{Bulge}$ & 7.113 $\pm$ 0.104 & -6.627 $\pm$ 0.311 & 15.305 $\pm$ 4.672 & 1.000 $\pm$ 0.003 & 25.217 $\pm$ 7.199 & 7.976 $\pm$ 1.109 & -3.117 $\pm$ 1.607 \\
		\hline
		\multirow{8}{*}{\shortstack{HB12 \\ Cen.}}
		& C & 3.782 $\pm$ 1.508 & -1.426 $\pm$ 2.577 & 1.083 $\pm$ 1.519 & 0.523 $\pm$ 0.386 & 6.756 $\pm$ 8.537 & 9.453 $\pm$ 2.815 & 3.031 $\pm$ 4.998 \\
		& Gini & 0.597 $\pm$ 0.217 & -0.376 $\pm$ 0.400 & 2.284 $\pm$ 10.540 & 0.652 $\pm$ 0.670 & 20.947 $\pm$ 10.220 & 6.367 $\pm$ 4.031 & -2.534 $\pm$ 5.301 \\
		& M$_{20}$ & -2.136 $\pm$ 0.740 & 0.377 $\pm$ 0.704 & 0.914 $\pm$ 4.920 & 0.492 $\pm$ 0.448 & 4.735 $\pm$ 8.175 & 9.387 $\pm$ 1.781 & -1.650 $\pm$ 4.983 \\
		& A & -0.005 $\pm$ 0.033 & 0.100 $\pm$ 0.034 & 14.173 $\pm$ 9.689 & -0.566 $\pm$ 0.633 & 3.520 $\pm$ 9.422 & 8.901 $\pm$ 2.708 & 3.695 $\pm$ 4.850 \\
		& GMB & -0.234 $\pm$ 0.391 & 0.641 $\pm$ 0.746 & 1.727 $\pm$ 1.137 & 0.882 $\pm$ 0.698 & 8.864 $\pm$ 8.085 & 11.461 $\pm$ 1.737 & 0.420 $\pm$ 4.784 \\
		& $r_{50}$ & 2.469 $\pm$ 0.086 & 8.465 $\pm$ 0.440 & 0.758 $\pm$ 0.030 & -0.381 $\pm$ 0.097 & 2.358 $\pm$ 0.387 & 11.997 $\pm$ 0.060 & -2.221 $\pm$ 0.444 \\
		& $n$ & 4.308 $\pm$ 0.037 & -0.980 $\pm$ 0.481 & 0.545 $\pm$ 0.212 & -0.744 $\pm$ 0.713 & 4.346 $\pm$ 10.962 & 11.169 $\pm$ 0.093 & -0.757 $\pm$ 2.803 \\
		& $n_{Bulge}$ & 6.761 $\pm$ 0.043 & -2.891 $\pm$ 2.635 & 0.531 $\pm$ 0.094 & -0.826 $\pm$ 0.687 & 4.514 $\pm$ 10.794 & 11.271 $\pm$ 0.037 & -2.821 $\pm$ 2.919 \\
		\hline
		\multirow{8}{*}{\shortstack{HB12 \\ Sat.}}
		& C & 2.812 $\pm$ 0.357 & 1.754 $\pm$ 0.952 & 1.779 $\pm$ 1.250 & 0.760 $\pm$ 0.435 & 4.981 $\pm$ 8.817 & 13.316 $\pm$ 1.192 & 3.720 $\pm$ 5.022 \\
		& Gini & 0.677 $\pm$ 0.200 & -0.178 $\pm$ 0.345 & 5.543 $\pm$ 10.342 & 0.913 $\pm$ 0.669 & 16.452 $\pm$ 10.191 & 6.249 $\pm$ 4.226 & -9.720 $\pm$ 5.215 \\
		& M$_{20}$ & -3.312 $\pm$ 0.704 & 1.737 $\pm$ 0.637 & 5.072 $\pm$ 3.103 & -0.204 $\pm$ 0.459 & 16.326 $\pm$ 8.097 & 7.131 $\pm$ 1.543 & 2.015 $\pm$ 5.097 \\
		& A & -0.003 $\pm$ 0.030 & 0.064 $\pm$ 0.031 & 1.203 $\pm$ 9.852 & -0.003 $\pm$ 0.643 & 18.142 $\pm$ 9.218 & 12.247 $\pm$ 2.703 & -0.487 $\pm$ 4.968 \\
		& GMB & -0.185 $\pm$ 0.098 & 0.580 $\pm$ 0.121 & 2.614 $\pm$ 1.120 & 0.936 $\pm$ 0.654 & 13.086 $\pm$ 8.353 & 12.052 $\pm$ 0.837 & 1.616 $\pm$ 4.892 \\
		& $r_{50}$ & 2.289 $\pm$ 0.058 & 6.355 $\pm$ 0.535 & 1.219 $\pm$ 0.132 & -0.604 $\pm$ 0.319 & 25.024 $\pm$ 7.175 & 11.991 $\pm$ 0.069 & 4.965 $\pm$ 2.956 \\
		& $n$ & 3.285 $\pm$ 0.034 & -2.028 $\pm$ 0.224 & 0.734 $\pm$ 0.837 & 0.847 $\pm$ 0.075 & 1.942 $\pm$ 2.534 & 8.917 $\pm$ 0.608 & -2.332 $\pm$ 1.665 \\
		& $n_{Bulge}$ & 7.153 $\pm$ 0.047 & -7.165 $\pm$ 0.322 & 0.434 $\pm$ 0.086 & 0.393 $\pm$ 0.304 & 4.170 $\pm$ 1.384 & 8.742 $\pm$ 0.048 & -0.785 $\pm$ 0.746 \\
		\hline
		\multirow{8}{*}{\shortstack{HB13 \\ Cen.}}
		& C & 3.819 $\pm$ 1.348 & -0.937 $\pm$ 2.327 & 1.000 $\pm$ 1.478 & 0.757 $\pm$ 0.480 & 8.887 $\pm$ 8.662 & 10.140 $\pm$ 2.723 & 2.644 $\pm$ 5.054 \\
		& Gini & 0.647 $\pm$ 0.242 & -0.477 $\pm$ 0.398 & 3.520 $\pm$ 10.127 & 0.434 $\pm$ 0.693 & 16.160 $\pm$ 9.862 & 4.195 $\pm$ 4.044 & -4.348 $\pm$ 5.119 \\
		& M$_{20}$ & -2.102 $\pm$ 0.834 & 0.340 $\pm$ 0.807 & 0.588 $\pm$ 9.420 & 0.299 $\pm$ 0.506 & 19.157 $\pm$ 9.097 & 9.741 $\pm$ 2.663 & -6.523 $\pm$ 4.883 \\
		& A & -0.015 $\pm$ 0.048 & 0.060 $\pm$ 0.050 & 2.445 $\pm$ 9.588 & 0.984 $\pm$ 0.644 & 9.508 $\pm$ 9.346 & 11.056 $\pm$ 2.666 & -2.074 $\pm$ 4.918 \\
		& GMB & -0.177 $\pm$ 0.389 & 0.751 $\pm$ 0.742 & 0.766 $\pm$ 1.275 & 0.142 $\pm$ 0.729 & 6.660 $\pm$ 8.418 & 11.232 $\pm$ 1.543 & -5.040 $\pm$ 4.753 \\
		& $r_{50}$ & 3.560 $\pm$ 0.011 & 15.416 $\pm$ 0.506 & 0.513 $\pm$ 0.008 & 0.044 $\pm$ 0.061 & 1.483 $\pm$ 0.118 & 12.000 $\pm$ 0.003 & -0.982 $\pm$ 0.349 \\
		& $n$ & 3.603 $\pm$ 0.037 & -2.015 $\pm$ 0.262 & 7.250 $\pm$ 2.241 & 0.999 $\pm$ 0.008 & 24.597 $\pm$ 8.452 & 8.657 $\pm$ 0.935 & -5.290 $\pm$ 3.407 \\
		& $n_{Bulge}$ & 7.561 $\pm$ 1.324 & -3.887 $\pm$ 1.390 & 0.713 $\pm$ 0.667 & -0.607 $\pm$ 0.851 & 5.767 $\pm$ 12.015 & 10.343 $\pm$ 0.139 & -0.020 $\pm$ 2.093 \\
		\hline
		\multirow{8}{*}{\shortstack{HB13 \\ Sat.}}
		& C & 2.892 $\pm$ 0.260 & 0.674 $\pm$ 1.064 & 0.802 $\pm$ 1.408 & 0.768 $\pm$ 0.342 & 3.199 $\pm$ 9.336 & 11.192 $\pm$ 1.600 & -0.413 $\pm$ 4.932 \\
		& Gini & 0.687 $\pm$ 0.198 & -0.172 $\pm$ 0.352 & 2.225 $\pm$ 10.687 & 0.632 $\pm$ 0.663 & 22.345 $\pm$ 10.411 & 8.855 $\pm$ 4.052 & -8.618 $\pm$ 5.211 \\
		& M$_{20}$ & -2.185 $\pm$ 0.681 & 0.529 $\pm$ 0.600 & 1.417 $\pm$ 2.250 & 0.115 $\pm$ 0.424 & 6.602 $\pm$ 8.710 & 9.085 $\pm$ 1.636 & 1.729 $\pm$ 4.581 \\
		& A & -0.064 $\pm$ 0.027 & 0.091 $\pm$ 0.029 & 1.778 $\pm$ 9.989 & -0.150 $\pm$ 0.657 & 16.832 $\pm$ 9.664 & 11.104 $\pm$ 2.659 & -9.763 $\pm$ 4.962 \\
		& GMB & -0.198 $\pm$ 0.082 & 0.659 $\pm$ 0.132 & 1.117 $\pm$ 0.619 & 0.747 $\pm$ 0.525 & 2.912 $\pm$ 8.620 & 11.712 $\pm$ 0.448 & 0.230 $\pm$ 4.800 \\
		& $r_{50}$ & 2.650 $\pm$ 0.036 & 7.479 $\pm$ 0.506 & 0.845 $\pm$ 0.039 & -0.745 $\pm$ 0.039 & 3.558 $\pm$ 0.479 & 11.999 $\pm$ 0.054 & -4.148 $\pm$ 0.450 \\
		& $n$ & 3.437 $\pm$ 0.043 & -1.772 $\pm$ 0.104 & 0.847 $\pm$ 0.212 & 0.821 $\pm$ 0.080 & 3.997 $\pm$ 0.890 & 9.663 $\pm$ 0.069 & -0.018 $\pm$ 0.254 \\
		& $n_{Bulge}$ & 6.403 $\pm$ 0.035 & -9.243 $\pm$ 2.147 & 1.521 $\pm$ 0.293 & -0.964 $\pm$ 0.022 & 4.499 $\pm$ 0.693 & 11.424 $\pm$ 0.484 & -9.953 $\pm$ 1.401 \\
		\hline
	\end{tabular}
\end{table*}
%\end{landscape}

\end{appendix}

\end{document}